\documentclass[10pt,prd,preprintnumbers,amsmath,amssymb,nofootinbib]{revtex4}
\usepackage{cancel}
\usepackage{multirow}
\usepackage{xcolor}
\usepackage{graphicx}
\usepackage{subfigure}
\usepackage[colorlinks,
            citecolor=blue,
            anchorcolor=red,
            menucolor=red,
            linkcolor=red,
            filecolor=red,
            runcolor=red,
            urlcolor=blue,
            frenchlinks=red]{hyperref}
\begin{document}
\title{Double-heavy tetraquark states with heavy diquark-antiquark symmetry}

\author{Jian-Bo Cheng$^1$, Shi-Yuan Li$^1$, Yan-Rui Liu$^1$, Zong-Guo Si$^1$, Tao Yao$^1$}
\affiliation{ $^1$School of Physics, Shandong University, Jinan, Shandong, 250100, China}

\date{\today}

\begin{abstract}

We calculate the masses of the $QQ\bar{q}\bar{q}$ ($Q=c,b$; $q=u,d,s$) tetraquark states with the aid of heavy diquark-antiquark symmetry (HDAS) and the chromomagnetic interaction (CMI) model. The masses of the highest-spin ($J=2$) tetraquarks that have only the $(QQ)_{\bar{3}_c}(\bar{q}\bar{q})_{3_c}$ color structure are related with those of conventional hadrons using HDAS. Thereafter, the masses of their partner states are determined with the mass splittings in the CMI model. Our numerical results reveal that: (i) the lightest $cc\bar{n}\bar{n}$ ($n=u,d$) is an $I(J^P)=0(1^+)$ state around 3929 MeV (53 MeV above the $DD^*$ threshold) and none of the double-charm tetraquarks are stable; (ii) the stable double-bottom tetraquarks are the lowest $0(1^+)$ $bb\bar{n}\bar{n}$ around 10488 MeV ($\approx116$ MeV below the $BB^*$ threshold) and the lowest $1/2(1^+)$ $bb\bar{n}\bar{s}$ around 10671 MeV ($\approx20$ MeV below the $BB_s^*/B_sB^*$ threshold); and (iii) the two lowest $bc\bar{n}\bar{n}$ tetraquarks, namely the lowest $0(0^+)$ around 7167 MeV and the lowest $0(1^+)$ around 7223 MeV, are near-threshold states. Moreover, we discuss the constraints on the masses of double-heavy hadrons. Specifically, for the lowest nonstrange tetraquarks, we obtain $T_{cc}<3965$ MeV, $T_{bb}<10627$ MeV, and $T_{bc}<7199$ MeV.

\end{abstract}

\maketitle

\section{Introduction}\label{sec:Introduction}
Recently, the LHCb Collaboration \cite{Aaij:2017ueg} observed the doubly charmed baryon $\Xi_{cc}^{++}$ in the $\Lambda_{c}^{+}K^{-}\pi^{+}\pi^{+}$ mass distribution. Its mass was determined to be $3621.40\pm0.72$(stat.)$\pm$0.27(syst.)$\pm$0.14($\Lambda_{c}^{+}$) MeV/c$^{2}$. This value is 100 MeV higher than the mass of $\Xi_{cc}^{+}$, which was determined in the channels $\Lambda_{c}^{+}K^{-}\pi^{+}$ and $pD^+K^-$ by the SELEX Collaboration \cite{Mattson:2002vu,Ocherashvili:2004hi} more than fifteen years ago. The doubly heavy baryons $\Xi_{cc}^{++}$ and $\Xi_{cc}^{+}$ have also been searched for in the FOCUS \cite{Ratti:2003ez}, BABAR \cite{Aubert:2006qw}, and Belle \cite{Kato:2013ynr} detectors, with negative results. Thus far, the LHCb Collaboration has still not been able to confirm the $\Xi_{cc}^{+}$ baryon \cite{Aaij:2019jfq}.

The confirmation of $\Xi_{cc}^{++}$ has important implications as it indicates that two identical charm quarks can exist in a hadronic state. The observation of this baryon has motivated several theoretical discussions regarding the possible double-charm tetraquark $T_{cc}$ and its partner states\footnote{We use $T_{cc}$ to specifically denote the lowest $cc\bar{u}\bar{d}$ tetraquark state with $I(J^P)=0(1^+)$ in this article. Similarly, $T_{bb}$ represents the lowest $bb\bar{u}\bar{d}$ with $I(J^P)=0(1^+)$. However, $T_{bc}$ represents the lowest $bc\bar{u}\bar{d}$ with $I(J^P)=0(0^+)$.}. In the literature, various approaches have been applied to the double-heavy tetraquark structures $QQ\bar{q}\bar{q}$ ($Q=c,b$; $q=u,d,s$), including the color-magnetic interaction (CMI) model \cite{Lee:2009rt,Hyodo:2012pm,Hyodo:2017hue,Luo:2017eub,Yan:2018gik}, quark-level models \cite{Lipkin:1986dw,Zouzou:1986qh,Semay:1994ht,Pepin:1996id,Brink:1998as,Feng:2013kea,Karliner:2017qjm,Park:2018wjk,Caramees:2018oue,Hernandez:2019eox,Yang:2009zzp,Yang:2019itm,Bedolla:2019zwg,Yu:2019sxx,Wallbott:2020jzh,Deng:2018kly,Tan:2020ldi,Lu:2020rog,Ebert:2007rn,Yang:2020fou}, QCD sum rule method \cite{Du:2012wp,Chen:2013aba,Wang:2017dtg,Agaev:2019kkz,Tang:2019nwv,Agaev:2019lwh,Navarra:2007yw,Gao:2020ogo}, lattice QCD simulation \cite{Brown:2012tm,Ikeda:2013vwa,Bicudo:2015vta,Francis:2016hui,Francis:2018jyb,Junnarkar:2018twb,Leskovec:2019ioa,Hudspith:2020tdf}, and holographic model \cite{Liu:2019yye}. One may consult Ref. \cite{Liu:2019zoy} for further discussions on such exotic states and related methods.

In Ref. \cite{Luo:2017eub}, the $QQ\bar{q}\bar{q}$ states were systematically studied using a CMI model in which the color mixing effects between $(QQ)_{\bar{3}_{c}}(\bar{q}\bar{q})_{3_{c}}$ and $(QQ)_{6_{c}}(\bar{q}\bar{q})_{\bar{6}_{c}}$ structures were considered, and the thresholds of the meson-meson channels were treated as reference scales to estimate the tetraquark masses. According to a series of studies on multiquark states that used the above model \cite{Luo:2017eub,Wu:2016gas,Chen:2016ont,Wu:2016vtq,Wu:2017weo,Zhou:2018pcv,Li:2018vhp,Wu:2018xdi,An:2019idk,Cheng:2019obk}, it appears that the method that uses thresholds usually yields underestimated masses \cite{Liu:2019zoy}. A possible reason for the underestimation is that the color-electric contribution to the two heavy quarks in the tetraquarks was not explicitly considered \cite{Karliner:2017qjm,Weng:2018mmf}. Considering the color-Coulomb interaction, the binding energy of two heavy quarks exhibits a positive correlation with their reduced mass. When the two heavy quarks are separated by a large distance, the $QQ\bar{q}\bar{q}$ state will form a mixed $(Q\bar{q})_{1c}(Q\bar{q})_{1c}$-$(Q\bar{q})_{8c}(Q\bar{q})_{8c}$ meson-meson type structure. In contrast, if the two heavy quarks move in a small spatial region because of the attraction, they may form a $\bar{3}_{c}$ substructure and the tetraquark can be treated as a diquark-antidiquark state. In this case, it is not necessary for the distance between the two light antiquarks to be small due to the considerable relativistic effect and small color-Coulomb potential. In this study, we aimed to perform a further investigation of double-heavy tetraquark systems, particularly the nonstrange double-charm system.

The heavy quark flavor-spin symmetry appears in the limit $m_Q\to\infty$, and it is used extensively to study the properties of heavy quark hadrons. For states containing two heavy quarks, the heavy diquark-antiquark symmetry (HDAS) can also be considered \cite{Carlson:1987hh, Lichtenberg:1989ix,Savage:1990di, Anselmino:1992vg,Brambilla:2005yk,Fleming:2005pd,Cohen:2006jg}. According to this symmetry, the mass splittings between $QQq$ baryons and those between $\bar{Q}'q$ mesons can be related with the correspondence $QQ\leftrightarrow \bar{Q}'$. The consideration is based on the observations that i) the size of $QQ$ is small in the heavy quark limit, ii) the color representations of $QQ$ and $\bar{Q}'$ are both $\bar{3}_c$, and iii) the interaction between light and heavy components is suppressed, even though $QQ$ and $\bar{Q}'$ have different spins. Similarly, one can relate the double-heavy tetraquarks $QQ\bar{q}\bar{q}$, with the color structure $ (QQ)_{3_c}$, to the singly heavy antibaryons $\bar{Q}'\bar{q}\bar{q}$. According to the HDAS, for example, we can estimate the mass of a double-charm tetraquark with the relation $T_{cc\bar{n}\bar{n}}-\Xi_{cc}^{*}=\Sigma_{b}^{*}-\bar{B}^{*}$, where $n=u,d$, and the hadron symbol represents its mass. Obviously, the required unknown input is only the mass of $\Xi_{cc}^{*}$, which can be estimated with the experimental mass of $\Xi_{cc}^{++}$ in the CMI model. Although the quantum numbers of LHCb $\Xi_{cc}^{++}$ have not been measured, its mass is very close to the theoretical value of the ground state predicted by Karliner and Rosner in Ref. \cite{Karliner:2014gca}. In the following calculations, we will use $m_{\Xi_{cc}}$=3621 MeV as the input to estimate the masses of the double-charm tetraquark states. Other double-heavy tetraquarks will also be systematically investigated. If the LHCb $\Xi_{cc}$ is actually the $\Xi_{cc}^*$ state with spin=3/2, $m_{\Xi_{cc}^*}-m_{\Xi_{cc}}\approx 70$ MeV should be subtracted from the obtained masses of relevant tetraquarks.

Unlike the conventional hadrons, the $QQ$ diquark in tetraquarks may also be in the color $6_c$ representation. In Ref. \cite{Hyodo:2012pm}, the mass of $I(J^P)=0(1^+)$ $T_{cc}$ with $6_c$ $cc$ was estimated. To include the mixing effects between the $(cc)_{\bar{3}_c}(\bar{n}\bar{n})_{3_c}$ and $(cc)_{6_c}(\bar{n}\bar{n})_{\bar{6}_c}$ configurations and to estimate the masses of all the $(cc\bar{n}\bar{n})$ states, in the current study, we first need to identify the position of the $(cc)_{\bar{3}_c}(\bar{n}\bar{n})_{3_c}$ state determined with HDAS. Observing that the $J=2$ tetraquark is the pure $(cc)_{\bar{3}_c}(\bar{n}\bar{n})_{3_c}$ state because of the constraint from the Pauli principle, we directly relate its mass to that of $\Xi_{cc}^*$. Thereafter, we determine the masses of the lower tetraquark states from the mass splittings within the CMI model. Other double-heavy tetraquark states will be studied similarly. This concept is contrary to the estimation strategy adopted in our recent works \cite{Wu:2016gas,Chen:2016ont,Wu:2017weo,Luo:2017eub,Zhou:2018pcv,Wu:2016vtq,Li:2018vhp,Wu:2018xdi,An:2019idk,Cheng:2019obk,Cheng:2020nho} where the multiquark masses were determined from lower mass scales.

The remainder of this paper is organized as follows. In Sec.\ref{sec:formalism}, we present the method and formalism for the study. Thereafter, we provide our analysis and numerical results for the $cc\bar{u}\bar{d}$ states in Sec. \ref{sec: Tcc spectra} and the predictions on their partners in Sec. \ref{sec: TQQ' spectra}. In Sec. \ref{sec5}, we discuss the constraints on the masses of the involved heavy quark hadrons. Finally, Sec. \ref{sec: Discussions and summary} presents our discussions and a summary.

\section{Model and method}\label{sec:formalism}

\subsection{CMI model}\label{subsec:One gluon exchange model}

For ground state hadrons, the mass splittings of different spin states with the same quark content are mainly determined by the color-spin (color-magnetic) interaction in the quark model \cite{DeRujula:1975qlm},
\begin{eqnarray}
H_{CM}=-\sum_{i<j}C_{ij}\vec{\lambda}_{i}\cdot\vec{\lambda}_{j}\vec{\sigma}_{i}\cdot\vec{\sigma}_{j}.
\end{eqnarray}
Here, $i(j)$ represents the $i$th ($j$th) quark component of the tetraquark state, $\vec{\lambda}_{i}$ ($\vec{\lambda}_{j}$) is the vector containing the eight Gell-Mann matrices for the $i$th ($j$th) quark component, and $\vec{\sigma}_i$ ($\vec{\sigma}_j$) is the vector containing the three Pauli matrices for the $i$th ($j$th) quark component. It should be noted that $\vec{\lambda}_{i}$ ($\vec{\lambda}_{j}$) should be replaced with $-\vec{\lambda}_{i}^{*}$ ($-\vec{\lambda}_{j}^{*}$) if the quark component is an antiquark. The effective coupling parameters $C_{ij}$, which actually depend on the systems, include effects from the spatial wave function and the constituent quark masses. Thus, the mass formula in the CMI model is
\begin{eqnarray}\label{modelhamiltonian}
{M}=\sum_{i}m_{i}+\langle H_{CM}\rangle,
\end{eqnarray}
where the effective mass of the $i$th quark $m_{i}$ includes the constituent quark mass and contributions from other terms such as the color-Coulomb interaction and color confinement. In the following calculations, we will adopt the values of the parameters presented in Table \ref{CMI-parameters}, which are determined from the masses of conventional hadrons.
\begin{table}[htbp]
\caption{Coupling parameters (units: MeV) extracted from conventional hadrons. The value of $C_{c\bar{b}}$ is estimated using the mass splitting in the Godfrey-Isgur model \cite{Godfrey:1985xj}. Approximations $C_{cc}=k C_{c\bar{c}}$, $C_{bb}=k C_{b\bar{b}}$, $C_{cb}=k C_{c\bar{b}}$, and $C_{s\bar{s}}=C_{ss}/k$ are adopted \cite{KerenZur:2007vp,Lipkin:1986dx}, where $k\equiv C_{nn}/C_{n\bar{n}}\approx2/3$. The effective quark masses determined from the masses of the ground hadrons \cite{Tanabashi:2018oca} are $m_n=361.8$ MeV, $m_s=542.4$ MeV, $m_c=1724.1$ MeV, and $m_b=5054.4$ MeV.}\label{CMI-parameters}
\begin{tabular}{c|c|c|c|c|c|c|c|c|c}\hline
$C_{nn}=18.3$&$C_{ns}=12.0$&$C_{nc}=4.0$&$C_{nb}=1.3$&$C_{ss}=5.7$&$C_{sc}=4.4$&$C_{sb}=0.9$&$C_{cc}=3.2$&$C_{bb}=1.8$&$C_{cb}=2.0$\\
$C_{n\bar{n}}=29.9$&$C_{n\bar{s}}=18.7$&$C_{n\bar{c}}=6.6$&$C_{n\bar{b}}=2.1$&$C_{s\bar{s}}=9.3$&$C_{s\bar{c}}=6.7$&$C_{s\bar{b}}=2.3$&$C_{c\bar{c}}=5.3$&$C_{b\bar{b}}=2.9$&$C_{c\bar{b}}=3.3$\\\hline
\end{tabular}
\end{table}

The CMI model can provide relatively reasonable predictions for the mass splittings for various hadronic systems, but it is not good enough to estimate hadron masses because the effective quark masses have large uncertainties. Ref. \cite{Luo:2017eub} presented two methods for estimating the double-heavy tetraquark masses: one employs the mass formula \eqref{modelhamiltonian} and the other uses the modified formula
\begin{eqnarray}\label{eq:ref}
{M}=(M_{threshold}-\langle H_{CM}\rangle_{threshold})+\langle H_{CM}\rangle.
\end{eqnarray}
The first method, which uses the parameters in Table \ref{CMI-parameters}, provides theoretical upper limits for the masses. The differences between these upper limits and the ``realistic'' masses would be very large for heavy quark multiquark states. This can be observed, for example, in the results for conventional hadrons and the $cs\bar{c}\bar{s}$ tetraquark states \cite{Wu:2016gas}. A possible means of remedying the deviations is to include a color-electric term appropriately in the Hamiltonian of the CMI model \cite{Karliner:2014gca,Weng:2019ynv}. The second method yields more reasonable results than the first one, but it suffers from the problem of selecting the reference scale. For example, the threshold of the $J/\psi\phi$ channel leads to a lower mass $X(4140)$ than that of $D_s^+D_s^-$ does \cite{Wu:2016gas}. If the state has a mixed structure of $(c\bar{s})_{8_c}(\bar{c}s)_{8_c}$ and $(c\bar{s})_{1_c}(\bar{c}s)_{1_c}$, where the separation between $c$ and $\bar{s}$ is small and the distance between $c\bar{s}$ and $\bar{c}s$ is large, the choice to use $D_s^{(*)+}D_s^{(*)-}$ as a reference system to estimate the tetraquark masses is more natural, even though the resulting tetraquark mass is problably still lower than the measured one. In the $cc\bar{n}\bar{n}$ case, the threshold that may be used is only for the $D^{(*)}D^{(*)}$ channel. However, when $cc$ can be considered as a diquark with a small spatial separation, using such a threshold as a reference scale appears to not be a good choice. The $T_{cc}$ mass (approximately 100 MeV below the $DD^*$ threshold) will probably also be underestimated.

In the second method, better choices for estimating the tetraquark masses than hadron-hadron thresholds should exist. When estimating the tetraquark masses of $Qq\bar{Q}\bar{q}$ and $q_1q_2\bar{q}_3\bar{q}_4$ in Refs. \cite{Wu:2018xdi,Cheng:2020nho}, we attempted to relate the reference scales to the $X(4140)$ mass. The obtained masses were higher than those with the meson-meson thresholds. In the current study, we examine the results with the aid of the heavy diquark-antiquark symmetry. At present, it is not clear which choice yields more realistic results. Hopefully, future measurements regarding the predicted tetraquark states can provide an answer to this. In the following discussions, to compare the results, we will refer to the methods corresponding to these three reference choices as the threshold, $X(4140)$, and HDAS approaches.

\subsection{Diquark-antiquark symmetry}\label{subsec: Diquark-antiquark symmetry}

A diquark is generally assumed to be a color-$\bar{3}$ correlated quark-quark subsystem of a bound or resonant state. It shares certain similar properties with an antiquark, and diquark-antiquark symmetry (DAS) may exist, even though the diquark and antiquark have different masses and spins, and their dynamics are not necessarily the same. This symmetry works for hadrons that contain a heavy quark with a mass that is much larger than $\Lambda_{QCD}$. In this case, the heavy quark can be treated as a static color source, and many properties are independent of the quark mass. For example, if we treat the $\bar{n}\bar{n}$ antidiquark as a heavier light quark $n^\prime$, a $cc\bar{n}\bar{n}$ state will become a $ccn^\prime$ state, which resembles a baryon structure. Subsequently, the mass difference between the $ccn^\prime$ tetraquark and the $ccn$ baryon is independent of the heavy quark mass in the heavy quark limit. This similarity based on DAS provides a method for estimating the masses of unknown states from those of known baryons. Likewise, as discussed by Savage and Wise in \cite{Savage:1990di}, if the $cc$ heavy diquark is treated as an antiquark $\bar{Q}^\prime$, the masses of various $\bar{Q}^\prime\bar{n}\bar{n}$ tetraquarks can be related to those of $Qnn$ baryons. As DAS is an approximate symmetry, whether or not the predictions based on it are correct needs to be tested experimentally. Before estimating the tetraquark masses, we provide further explanations of the diquark-antiquark symmetry.

\begin{figure}[htbp]
\includegraphics[width=330pt]{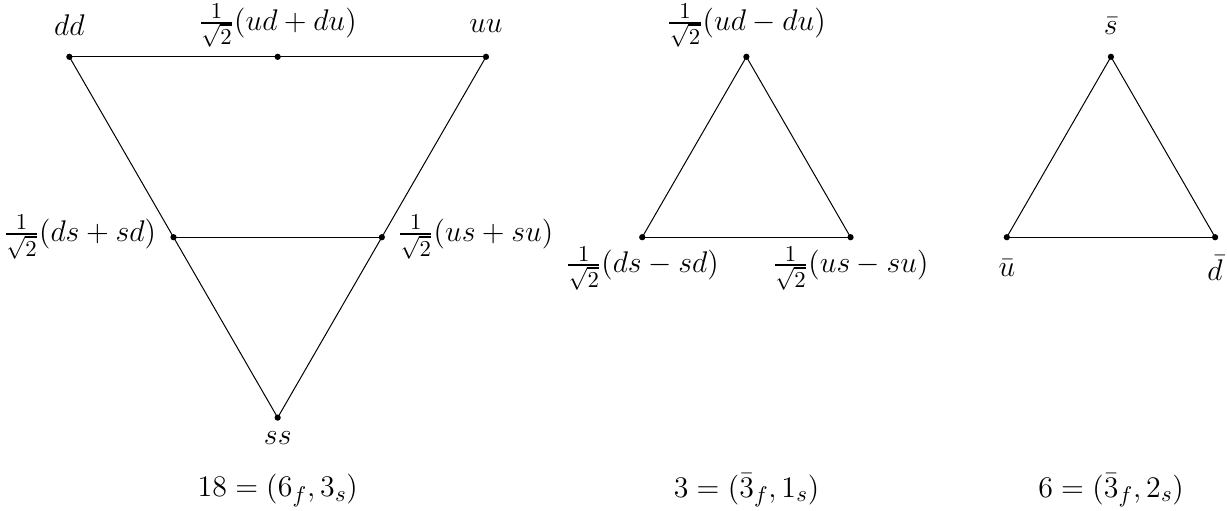}
\caption{$27$-multiplet in $SU(6)$ group with diquark-antiquark symmetry. The first number in the parentheses denotes the flavor $SU(3)$ representation and the second the spin $SU(2)$ representation.}\label{fig:27-multiplet}
\end{figure}

 It is known that there are three types of light quarks ($u$, $d$, and $s$) which form the base representation of flavor $SU(3)$. As quarks have spin, the flavor-spin $SU(6)$ can conventionally be used to classify various quark states. If a light diquark is a stable object, it has been argued that a symmetry exists between the baryons and mesons (diquarks and antiquarks) \cite{Miyazawa:1966mfa,Miyazawa:1968zz,Gao:1982cy}. Considering the color-$\bar{3}$ diquark and antiquark together, we obtain a flavor-spin $27$-plet. Its $SU(3)\otimes SU(2)$ decomposition reads $27=(6_{f},3_{s})+(\bar{3}_f,1_{s})+(\bar{3}_{f},2_{s})$. The substructure $(6_{f},3_{s})$ represents an $18$-plet with flavor-symmetric and spin-symmetric diquarks. Similarly, $(\bar{3}_{f},1_{s})$ represents a triplet with flavour-antisymmetric and spin-antisymmetric diquarks. The last substructure $(\bar{3}_{f},2_{s})$ represents the antiquark sextet. The decomposition indicates that the diquark and antiquark can be combined into the $SU(6/21)$ symmetry algebra. We present the group structures in Fig. \ref{fig:27-multiplet}.

Lattice QCD simulations have indicated such a diquark-antiquark symmetry \cite{Thacker:1987aq,Duke:1986sz}: the static quark-diquark potential is almost equal to the static quark-antiquark potential, and a quark-antiquark pair and a quark-diquark pair
have similar wave functions. However, according to Ref. \cite{Anselmino:1992vg}, the diquark-antiquark symmetry in light quark sector is broken for at least three reasons: (1) a diquark and an antiquark have different masses, lead to kinematical differences; (2) the diquark and antiquark have different spin-dependent and velocity-dependent terms; and (3) the diquark is not a point particle, and its finite size must affect its interactions. Thus, the breaking effects for the diquark-antiquark symmetry between the light diquarks and the light antiquarks are significant.

In contrast, for hadrons containing one heavy quark, the aforementioned symmetry breaking effects will be largely suppressed \cite{Anselmino:1992vg}. According to the heavy quark effective theory (HQET) \cite{Manohar:2000dt}, the kinematic and spin-dependent terms are inversely proportional to the heavy quark mass and make small contributions to the hadron mass. The size of the diquark is not a major concern, because the constituent quark model is still successful in handling the properties of conventional hadrons, even though the constituent quark has a comparable size to the diquark \cite{Anselmino:1992vg}. Thus, the size of diquark has no significant impact on the DAS.
Therefore, in general, there is possibly a better diquark-antiquark symmetry for hadrons containing one heavy quark. In Ref. \cite{Lichtenberg:1995kg}, Lichtenberg, Roncaglia, and Predazzi analyzed relations for the masses of heavy quark hadrons using the Feynman-Hellmann theorem and semiempirical formulas. They obtained some mass sum rules for heavy quark hadrons. We have listed a selection of these as follows:
\begin{eqnarray}
  D_{s}^{*}-D^{*} &=& \bar{B}_{s}^{*}-\bar{B}^{*}, \label{vec eq 1} \\
  \Sigma_{b}^{*}-\Sigma_{c}^{*} &=& \bar{B}^{*}-D^{*}, \label{vec eq 2} \\
  \Xi_{b}^{*}-\Xi_{c}^{*}&=& \bar{B}^{*}-D^{*}, \label{vec eq 3}\\
  \Omega_{b}^{*}-\Omega_{c}^{*} &=& \bar{B}^{*}-D^{*}, \label{vec eq 4}
\end{eqnarray}
which are adopted in the following discussions. These also follow from the heavy quark flavor symmetry. These relations can be confirmed by observing that the values of $(l.h.s.-r.h.s)$ are $12.9$ MeV, $-0.6$ MeV, $-9.5$ MeV, and $-21.5$ MeV (${\Omega_b^*}-{\Omega_b}=14.4$ MeV is used in the CMI model), respectively. The common feature of these four relations is that only highest spins are involved. In fact, the final three relations also satisfy the diquark-antiquark symmetry for light quarks in which the diquark spin is 1. It is true that a better DAS exists for hadrons containing one heavy quark than for those without heavy quarks. Although more relations can be found in Ref. \cite{Lichtenberg:1995kg}, it is not necessary to consider them in this work.

Our strategy for estimating the masses of double-heavy tetraquarks is to combine HDAS and the aforementioned four mass sum rules. To illustrate the concept, we temporarily focus only on Eq. \eqref{vec eq 1}. If we consider the heavy diquark-antiquark symmetry for the $cc$ diquark, we obtain
\begin{eqnarray}
\Omega_{cc}^{*}-\Xi_{cc}^{*}=\bar{B}_{s}^{*}-\bar{B}^{*}(=D_s^*-D^*), \label{vec eq 5}
\end{eqnarray}
which can be used to estimate the mass of $\Omega_{cc}^{*}$ with that of $\Xi_{cc}^{*}$. As a better heavy quark symmetry exists in bottom systems than in charmed systems, we use the masses of the bottom mesons. Here and in the following discussions, we only consider the highest spin hadrons while adopting the diquark-antiquark symmetry, because then, the possible contributions from other color or spin structures will be avoided. As explained in Sec. \ref{sec:Introduction}, we assume that the spin of the LHCb $\Xi_{cc}$ is 1/2. We can evaluate the mass of $\Xi_{cc}^*$ with the CMI model $\Xi_{cc}^{*}=\Xi_{cc}+16C_{cn}=3685$ MeV. Thereafter, the mass of $\Omega_{cc}^*$, namely $3776$ MeV, is obtained. By using the CMI model again, we can further obtain ${\Omega_{cc}}=3706$ MeV.

At present, doubly heavy baryons other than $\Xi_{cc}$ have not been observed, and the accuracy of Eq. \eqref{vec eq 5} cannot be verified. However, in theory, the reasonability of treating a $QQ$ diquark as a heavy antiquark $\bar{Q}^{\prime}$ can be argued. Because the heavy diquark has lower kinematic energy and less spin-dependent interaction, the heavy quark approximation works better than the single $Q$ case. Furthermore, the heavy diquark (with a light quark spectator) actually has better symmetry properties than the light diquark (with a heavy quark spectator) because a heavy diquark has small spatial separation, which means that its interaction with light quarks is not affected significantly by its size. In the following parts of this paper, we first focus on the double-charm tetraquark $cc\bar{n}\bar{n}$ states by treating them as systems that are composed of a small-size double-charm diquark and a light antidiquark. According to HDAS, the study of $cc\bar{n}\bar{n}$ states becomes that of heavy $Q^\prime nn$ ``baryons''. This approximate symmetry, together with the above mass sum rules, is evidently convenient for us to relate $cc\bar{n}\bar{n}$ to $\Xi_{cc}$. Focusing on the $cc\bar{n}\bar{n}$ states, we perform similar studies on other double-heavy tetraquarks. Whether or not the adopted approximation is effective should be tested in future experiments.

\section{$cc\bar{n}\bar{n}$ spectrum in terms of known $\Xi_{cc}$}\label{sec: Tcc spectra}

Based on the symmetry consideration, we obtain a good flavor-spin supermultiplet \cite{Lichtenberg:1989ix} that contains both tetraquark mesons $cc\bar{q}\bar{q}$ and three-quark baryons $ccq$. This double-charm supermultiplet is classified into three types of states,
\begin{eqnarray}
  6_{fs}&:& ccu, \ \ \ ccd, \ \ \ ccs; \label{cc-q} \\
  18_{fs}&:& cc\bar{u}\bar{u}, \ \ \ cc\bar{u}\bar{d}, \ \ \ cc\bar{d}\bar{d}, \ \ \ cc\bar{u}\bar{s}, \ \ \ cc\bar{d}\bar{s}, \ \ \ cc\bar{s}\bar{s} \label{cc-NN 1};\\
  \bar{3}_{fs}&:& cc\bar{u}\bar{d}, \ \ \ cc\bar{u}\bar{s}, \ \ \ cc\bar{d}\bar{s}. \label{cc-NN 2}
\end{eqnarray}
The members in Eq. \eqref{cc-NN 1} contain a light antidiquark with spin 1, and those in \eqref{cc-NN 2} have a light antidiquark with spin 0. We treat the $cc$ diquark as a heavy $\bar{3}_c$ ``antiquark'' $\bar{Q}^\prime$. Following the replacement, $cc\rightarrow \bar{Q}^\prime$, the $\Xi_{cc}^{*}$ with spin 3/2 and the $cc\bar{n}\bar{n}$ with spin 2 are transformed into $\bar{M}_{Q^\prime}^{*}=(\bar{Q}^\prime n)^{J=3/2}$ and $\bar{\Sigma}_{Q^\prime}^{*}=(\bar{Q}^\prime\bar{n}\bar{n})^{J=2}$, respectively, i.e.
\begin{eqnarray}
  &&\bar{\Xi}_{cc}^{*}(\bar{c}\bar{c}\bar{n})\rightarrow M_{Q^\prime}^{*}(Q^\prime\bar{n}),\label{cc-Q 1} \\
  &&{T}_{\bar{c}\bar{c}nn}^{I=1,J=2} \rightarrow  \Sigma_{Q^\prime}^{*}(Q^\prime nn). \label{cc-Q 2}
\end{eqnarray}
Note that only the highest spins are involved, and the $cc$ diquark has unique quantum numbers $color=\bar{3}_c,spin=1$. The forms on the right hand side remind us of the relation in Eq. \eqref{vec eq 2}, where the light diquark-antiquark symmetry is used. With that equation, one can naturally obtain
\begin{eqnarray}
T_{cc\bar{n}\bar{n}}^{I=1,J=2}-\Xi_{cc}^{*}=\Sigma_{c}^{*}-D^{*}=\Sigma_{b}^{*}-\bar{B}^{*}.\label{susyrelation-nn}
\end{eqnarray}
The obtained mass of $cc\bar{n}\bar{n}$ with $I=1,J=2$ is, thus, 4195 MeV (with $\Sigma_c^*$ and $D^*$) or 4194 MeV (with $\Sigma_b^*$ and $\bar{B}^{*}$). We select the later value owing to the better heavy quark symmetry for the bottom hadrons. Subsequently, we have a good reference hadron $T_{cc\bar{n}\bar{n}}^{I=1, J=2}$ and can estimate the masses of other $cc\bar{n}\bar{n}$ states by considering the CMI differences. When only $\bar{3}_c$ $cc$ is considered, the masses of the other three double-charm tetraquarks are $T_{cc\bar{n}\bar{n}}^{I=1,J=0}=4087$ MeV, $T_{cc\bar{n}\bar{n}}^{I=1,J=1}=4122$ MeV, and $T_{cc\bar{n}\bar{n}}^{I=0,J=1}=3961$ MeV. It is obvious that these tetraquarks are all above the $DD^*$ threshold (3876 MeV).

\begin{table}[!h]
\caption{Color-spin bases for $cc\bar{q}\bar{q}$ ($q=u,d,s$) states \cite{Luo:2017eub}. The superscripts indicate the spin, and the subscripts indicate the color representations.}\label{table:Tccbases}
\begin{tabular}{c|cccc}\hline\hline
States&$I(J^P)$&&Bases&\\
\hline
$(cc\bar{n}\bar{n})$
&$1(2^{+})$&$[(cc)_{\bar{3}}^{1}(\bar{n}\bar{n})_{3}^{1}]^{2}$&&\\
&$1(1^{+})$&$[(cc)_{\bar{3}}^{1}(\bar{n}\bar{n})_{3}^{1}]^{1}$&&\\
&$1(0^{+})$&$[(cc)_{\bar{3}}^{1}(\bar{n}\bar{n})_{3}^{1}]^{0}$&$[(cc)_{6}^{0}(\bar{n}\bar{n})_{\bar{6}}^{0}]^{0}$&\\
&$0(1^{+})$&$[(cc)_{\bar{3}}^{1}(\bar{n}\bar{n})_{3}^{0}]^{1}$&$[(cc)_{6}^{0}(\bar{n}\bar{n})_{\bar{6}}^{1}]^{1}$&\\
\hline\hline
$(cc\bar{n}\bar{s})$
&$(2^{+})$&$[(cc)_{\bar{3}}^{1}(\bar{n}\bar{s})_{3}^{1}]^{2}$&&\\
&$(1^{+})$&$[(cc)_{\bar{3}}^{1}(\bar{n}\bar{s})_{3}^{1}]^{1}$&$[(cc)_{\bar{3}}^{1}(\bar{n}\bar{s})_{3}^{0}]^{1}$&$[(cc)_{6}^{0}(\bar{n}\bar{s})_{\bar{6}}^{1}]^{1}$\\
&$(0^{+})$&$[(cc)_{\bar{3}}^{1}(\bar{n}\bar{s})_{3}^{1}]^{0}$&$[(cc)_{6}^{0}(\bar{n}\bar{s})_{\bar{6}}^{0}]^{0}$&\\
\hline\hline
$(cc\bar{s}\bar{s})$
&$(2^{+})$&$[(cc)_{\bar{3}}^{1}(\bar{s}\bar{s})_{3}^{1}]^{2}$&&\\
&$(1^{+})$&$[(cc)_{\bar{3}}^{1}(\bar{s}\bar{s})_{3}^{1}]^{1}$&&\\
&$(0^{+})$&$[(cc)_{\bar{3}}^{1}(\bar{s}\bar{s})_{3}^{1}]^{0}$&$[(cc)_{6}^{0}(\bar{s}\bar{s})_{\bar{6}}^{0}]^{0}$&\\
\hline\hline
\end{tabular}
\end{table}

\begin{figure}[!th]
\subfigure[$\,\,cc\bar{n}\bar{n}$]{ \label{fig: ccnn}
\includegraphics[width=130pt]{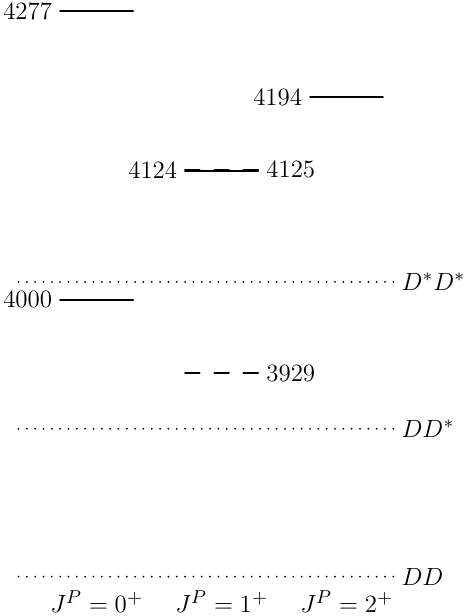}}\hspace{20pt}
\subfigure[$\,\,cc\bar{n}\bar{s}$]{ \label{fig: ccns}
\includegraphics[width=130pt]{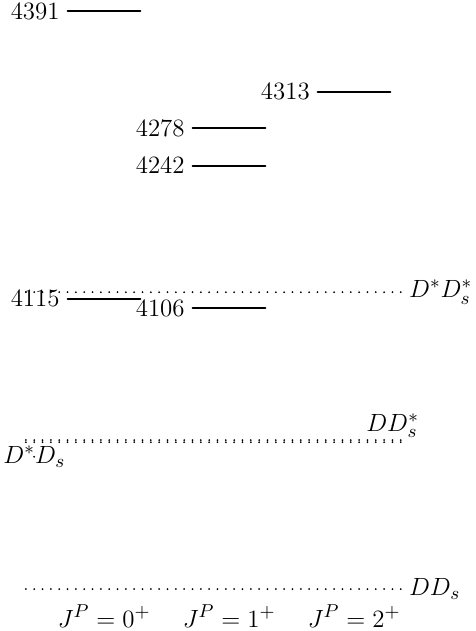}}\hspace{20pt}
\subfigure[$\,\,cc\bar{s}\bar{s}$]{ \label{fig: ccss}
\includegraphics[width=130pt]{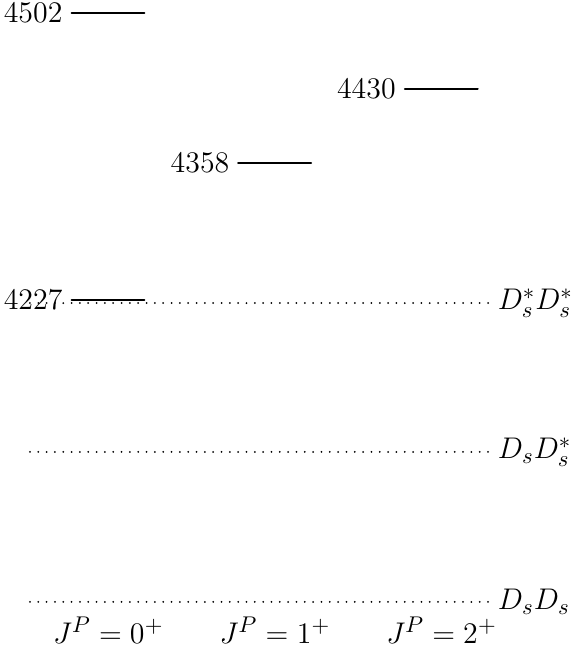}}
\caption{Relative positions of the double-charm tetraquark states (solid and dashed lines) and relevant meson-meson thresholds (dotted lines). The masses are given in MeV. In (a), the solid (dashed) lines denote the $I=1$ ($I=0$) states, and the almost degenerate masses 4124 MeV and 4125 MeV correspond to the isovector and isoscalar states, respectively.}\label{fig:cc}
\end{figure}

Although the color structure of $\Xi_{cc}$ is unique, that of the exotic $cc\bar{n}\bar{n}$ states is not. The mixing or channel coupling effects from the $(cc)_{6_c}(\bar{n}\bar{n})_{\bar{6}_c}$ color structure may significantly change the tetraquark masses. When such contributions are considered, two more tetraquarks appear. We collect all the color-spin bases \cite{Luo:2017eub} for the $cc\bar{n}\bar{n}$ states, which are displayed in Table \ref{table:Tccbases}. With these wave functions, we finally obtain the numerical results listed in the fourth column of Table \ref{table:Tcc coupled channel} and illustrated in Fig. \ref{fig: ccnn}. Moreover, we provide the masses estimated using Eq. \eqref{eq:ref} ($D^{(*)}D^{(*)}$ as the reference state) and those using Eq. \eqref{modelhamiltonian} in the table. They can be viewed as the theoretical lower (fifth column) and upper (sixth column) limits, respectively, in the present framework. Comparing these three results from different considerations, it is found that the new masses fall within the range constrained by the lower and upper limits, and the values are slightly larger than the averages of the two limits. The masses of the lowest $1(0^{+})$ and $0(1^{+})$ states in Table \ref{table:Tcc coupled channel} are evidently smaller than the above values without the $(cc)_{6_c}(\bar{n}\bar{n})_{\bar{6}_c}$ contributions. However, the $T_{cc}$ state, the mass of which in the HDAS approach is 155 MeV higher than the lower limit (3774 MeV), is still above the $DD^{*}$ threshold. The other $cc\bar{n}\bar{n}$ states are also above the respective fall-apart thresholds. Therefore, the HDAS approach results in an unstable $T_{cc}$, which is consistent with the conclusion obtained in Refs. \cite{Karliner:2017qjm,Eichten:2017ffp,Carlson:1987hh,Du:2012wp,Francis:2016hui,Francis:2018jyb,Lu:2020rog,Braaten:2020nwp,Park:2018wjk,Navarra:2007yw,Ebert:2007rn}.

\begin{table}[htbp]
\caption{Results for the $cc\bar{n}\bar{n}$ ($n=u,d$) states (unit: MeV). The second and third columns present the numerical values of the CMI matrices and their eigenvalues, respectively. The fifth and sixth columns list the masses estimated using Eq. \eqref{eq:ref} ($D^{(*)}D^{(*)}$ as the reference state) and Eq. \eqref{modelhamiltonian} (parameters presented in Table \ref{CMI-parameters}), respectively. These can be viewed as the theoretical lower limits (low.) and upper limits (up.), respectively, for the tetraquark masses in the current framework. The fourth column displays our predictions with the heavy diquark-antiquark symmetry (HDAS) consideration.}\label{table:Tcc coupled channel}
\begin{tabular}{c|ccccc}\hline
$(cc\bar{n}\bar{n})$&$\langle H_{CM}\rangle$&Eigenvalues&Mass (our)&Mass (low.)&Mass (up.)\\
$I(J^P)$&&&HDAS&$D^{(*)}D^{(*)}$&Eq. \eqref{modelhamiltonian}\\
\hline
$1(2^{+})$&$\left(\begin{array}{c}92.7\\\end{array}\right)$&$\left(\begin{array}{c}92.7\\\end{array}\right)$&$\left(\begin{array}{c}4194.4\\\end{array}\right)$&$\left(\begin{array}{c}4038.8\\\end{array}\right)$&$\left(\begin{array}{c}4264.5\\\end{array}\right)$\\
$1(1^{+})$&$\left(\begin{array}{c}22.3\\\end{array}\right)$&$\left(\begin{array}{c}22.3\\\end{array}\right)$&$\left(\begin{array}{c}4124.0\\\end{array}\right)$&$\left(\begin{array}{c}3968.4\\\end{array}\right)$&$\left(\begin{array}{c}4194.1\\\end{array}\right)$\\
$1(0^{+})$&$\left(\begin{array}{cc}-12.9&129.3\\129.3&86.2\\\end{array}\right)$&$\left(\begin{array}{c}-101.9\\175.1\\\end{array}\right)$&$\left(\begin{array}{c}3999.8\\4276.8\\\end{array}\right)$&$\left(\begin{array}{c}3844.3\\4121.3\\\end{array}\right)$&$\left(\begin{array}{c}4069.9\\4346.9\\\end{array}\right)$\\
$0(1^{+})$&$\left(\begin{array}{cc}-137.7&-74.7\\-74.7&-11.4\\\end{array}\right)$&$\left(\begin{array}{c}-172.4\\23.2\\\end{array}\right)$&$\left(\begin{array}{c}3929.3\\4124.9\\\end{array}\right)$&$\left(\begin{array}{c}3773.8\\3969.4\\\end{array}\right)$&$\left(\begin{array}{c}3999.4\\4195.0\\\end{array}\right)$\\
\hline
\end{tabular}
\end{table}

\section{Predictions for other double-heavy tetraquark states}\label{sec: TQQ' spectra}

We have obtained the $cc\bar{n}\bar{n}$ spectrum with the mass of the LHCb $\Xi_{cc}$ state by considering the diquark-antiquark symmetry in the CMI model. It is natural to extend the study to other double-heavy tetraquark states, including $cc\bar{n}\bar{s}$, $cc\bar{s}\bar{s}$, $bb\bar{n}\bar{n}$, $bb\bar{n}\bar{s}$, $bb\bar{s}\bar{s}$, $bc\bar{n}\bar{n}$, $bc\bar{n}\bar{s}$, and $bc\bar{s}\bar{s}$. In Ref. \cite{Cheng:2020nho}, we estimated the masses of $bc\bar{n}\bar{n}$ and $bc\bar{n}\bar{s}$ states by using $X(4140)$ as a reference system. It will be instructive to compare results when using different reference states.

\subsection{$cc\bar{n}\bar{s}$ and $cc\bar{s}\bar{s}$ states}\label{subsec: Tcc stranged}

By following a similar procedure to that in Eq. \eqref{susyrelation-nn}, we can easily obtain two relations from Eqs. \eqref{vec eq 3} and \eqref{vec eq 4}, respectively,
\begin{eqnarray}
T_{cc\bar{n}\bar{s}}^{J=2}-\Xi_{cc}^* &=& \Xi_{c}^{*}-D^{*}=\Xi_{b}^{*}-\bar{B}^{*},\label{susyrelation-ns}\\
T_{cc\bar{s}\bar{s}}^{J=2}-\Omega_{cc}^* &=& \Omega_{c}^{*}-D_{s}^{*}=\Omega_{b}^{*}-\bar{B}_{s}^{*}.\label{susyrelation-ss}
\end{eqnarray}
As the heavy quark symmetry breaking effects are larger for charmed systems than for bottom systems, we use the masses of $\Xi_b^*$ and $\bar{B}^{*}$ in the former relation. To obtain the latter relation, we employ Eq. \eqref{vec eq 1} so that no $u$ or $d$ quark is involved. Actually, if the heavy meson is sufficiently heavy, the difference between $u$, $d$, and $s$ cases can be neglected. 
Because the mass of $\Omega_b^*$ has not yet been measured, we opt to use the masses of $\Omega_c^*$ and $D_s^*$. With $\Xi_{cc}^*=3685$ MeV and $\Omega_{cc}^*=3776$ MeV, estimated using Eq. \eqref{vec eq 5}, we obtain $T_{cc\bar{n}\bar{s}}^{J=2}=4313$ MeV and $T_{cc\bar{s}\bar{s}}^{J=2}=4430$ MeV. These will be treated as reference scales to determine the masses of other double-charm strange tetraquarks.

Before proceeding further, we investigate the masses of $T_{cc\bar{n}\bar{n}}^{J=2}$, $T_{cc\bar{n}\bar{s}}^{J=2}$, and $T_{cc\bar{s}\bar{s}}^{J=2}$ in different approaches. The current estimation yields 177 MeV, 193 MeV, and 205 MeV for the mass distances measured from the $D^*D^*$, $D^*D_s^*$, and $D_s^*D_s^*$ thresholds, respectively, which are gradually increasing numbers. Those in Ref. \cite{Luo:2017eub} are gradually decreasing numbers, 23 MeV, 8 MeV, and $-6$ MeV. Therefore, no stable $J=2$ tetraquarks are obtained in this study, whereas the states in Ref. \cite{Luo:2017eub} are around their fall-apart thresholds. This feature is an apparent difference between the HDAS approach and the threshold approach.

With the above reference states, $T_{cc\bar{n}\bar{s}}^{J=2}$ and $T_{cc\bar{s}\bar{s}}^{J=2}$, and the mass splittings in the CMI model, we can estimate the masses of strange partners of $cc\bar{n}\bar{n}$ states. The base structures for the calculation are presented in Table \ref{table:Tccbases}. We list the numerical results for all of the $(cc\bar{n}\bar{s})$ and $(cc\bar{s}\bar{s})$ states in Table \ref{table:Tccpartners}, where we also display the theoretical lower and upper limits for the tetraquark masses. Comparing the values in 4th, 5th, and 6th columns, it is obvious that our results with diquark-antiquark symmetry are slightly larger than the averages of the two limits, which is the same feature as in the $cc\bar{n}\bar{n}$ case. The relative positions for the $(cc\bar{n}\bar{s})$ and $(cc\bar{s}\bar{s})$ tetraquark states are illustrated in Fig. \ref{fig: ccns} and Fig. \ref{fig: ccss}, respectively. According to the figure, similar to the $cc\bar{n}\bar{n}$ case, all the obtained doule-charm strange tetraquarks can decay through rearrangement mechanisms, and no such stable states exist. This observation is different from that in Ref. \cite{Luo:2017eub}, where the lowest $1^+$ $cc\bar{n}\bar{s}$ is stable.

\begin{table}[htbp]
\caption{Results for $cc\bar{n}\bar{s}$ and $cc\bar{s}\bar{s}$ states (unit: MeV). The second and third columns provide the numerical values of the CMI matrices and their eigenvalues, respectively. The fifth and sixth columns list the masses estimated using Eq. \eqref{eq:ref} (with $D^{(*)}D_{s}^{(*)}/D_{s}^{(*)}D_{s}^{(*)}$ as the reference state) and Eq. \eqref{modelhamiltonian} (parameters provided in Table \ref{CMI-parameters}), respectively. These can be viewed as the theoretical lower limits (low.) and upper limits (up.), respectively, for the tetraquark masses in the current framework. The fourth column displays our predictions with the heavy diquark-antiquark symmetry (HDAS) consideration.}\label{table:Tccpartners}
\begin{tabular}{c|ccccc}\hline
$(cc\bar{n}\bar{s})$&$\langle H_{CM}\rangle$&Eigenvalues&Mass (our)&Mass (low.)&Mass(up.)\\
$J^{P}$&&&HDAS&$D^{(*)}D_{s}^{(*)}$&Eq. \eqref{modelhamiltonian}\\
\hline
$2^{+}$&$\left(\begin{array}{c}76.1\\\end{array}\right)$&$\left(\begin{array}{c}76.1\\\end{array}\right)$&$\left(\begin{array}{c}4313.3\\\end{array}\right)$&$\left(\begin{array}{c}4125.4\\\end{array}\right)$&$\left(\begin{array}{c}4428.5\\\end{array}\right)$\\
$1^{+}$&$\left(\begin{array}{ccc}5.2&0.0&0.0\\0.0&-87.3&-75.2\\0.0&-75.2&-3.0\\\end{array}\right)$&$\left(\begin{array}{c}-131.4\\5.2\\41.1\\\end{array}\right)$&$\left(\begin{array}{c}4105.8\\4242.4\\4278.3\\\end{array}\right)$&$\left(\begin{array}{c}3917.9\\4054.5\\4090.4\\\end{array}\right)$&$\left(\begin{array}{c}4221.0\\4357.6\\4393.5\\\end{array}\right)$\\
$0^{+}$&$\left(\begin{array}{cc}-30.3&130.3\\130.3&61.0\\\end{array}\right)$&$\left(\begin{array}{c}-122.7\\153.4\\\end{array}\right)$&$\left(\begin{array}{c}4114.5\\4390.6\\\end{array}\right)$&$\left(\begin{array}{c}3926.6\\4202.7\\\end{array}\right)$&$\left(\begin{array}{c}4229.7\\4505.8\\\end{array}\right)$\\
\hline\hline
($cc\bar{s}\bar{s}$)&$\langle H_{CM}\rangle$&Eigenvalues&Mass(our)&Mass(low.)&Mass(up.)\\
$J^{P}$&&&HDAS&$D_{s}^{(*)}D_{s}^{(*)}$&Eq. \eqref{modelhamiltonian}\\
\hline
$2^{+}$&$\left(\begin{array}{c}59.6\\\end{array}\right)$&$\left(\begin{array}{c}59.6\\\end{array}\right)$&$\left(\begin{array}{c}4429.9\\\end{array}\right)$&$\left(\begin{array}{c}4211.6\\\end{array}\right)$&$\left(\begin{array}{c}4592.6\\\end{array}\right)$\\
$1^{+}$&$\left(\begin{array}{c}-11.9\\\end{array}\right)$&$\left(\begin{array}{c}-11.9\\\end{array}\right)$&$\left(\begin{array}{c}4358.4\\\end{array}\right)$&$\left(\begin{array}{c}4140.1\\\end{array}\right)$&$\left(\begin{array}{c}4521.1\\\end{array}\right)$\\
$0^{+}$&$\left(\begin{array}{cc}-47.6&131.3\\131.3&35.8\\\end{array}\right)$&$\left(\begin{array}{c}-143.7\\131.8\\\end{array}\right)$&$\left(\begin{array}{c}4226.6\\4502.1\\\end{array}\right)$&$\left(\begin{array}{c}4008.3\\4283.8\\\end{array}\right)$&$\left(\begin{array}{c}4389.3\\4664.8\\\end{array}\right)$\\
\hline
\end{tabular}
\end{table}

\subsection{$bb\bar{q}\bar{q}$ and $bc\bar{q}\bar{q}$ states}\label{subsec: Spectrum of Tbb and Tbc system}

According to the diquark-antiquark symmetry, no stable double-charm tetraquark states exist. In the bottom case, the attractive color-Coulomb interaction between the two heavy quarks may be strong enough to aid in the formation of stable tetraquarks. Next, we investigate the $bb\bar{q}\bar{q}$ and $bc\bar{q}\bar{q}$ systems, where $q=u$, $d$, or $s$.
First, we focus on the double-bottom tetraquarks, which have exactly the same group structure as the double-charm states. For the HDAS relations and the wave function bases, we simply need to perform a simple substitution of $(bb)$ for $(cc)$ in Eqs. \eqref{susyrelation-nn}-\eqref{susyrelation-ss} and Table \ref{table:Tccbases}. However, considerable difficulty arises in applying the formulas, as the masses of $\Xi_{bb}^*$ and $\Omega_{bb}^*$ have not been measured. In this study, we need to select appropriate predictions for their values from various investigations.

\begin{table}[htbp]
\caption{Theoretical predictions for the masses of doubly heavy baryons (unit: MeV) in various approaches: lattice QCD \cite{Brown:2014ena}, chromomagnetic models \cite{Weng:2018mmf,Karliner:2014gca,Karliner:2018hos}, relativistic quark model \cite{Ebert:2002ig}, nonrelativistic quark model \cite{Roberts:2007ni,Albertus:2006ya,SilvestreBrac:1996bg}, bag model \cite{He:2004px}, Bethe-Salpeter equation \cite{Weng:2010rb}, and QCD sum rules \cite{Wang:2010hs,Wang:2010vn}.}\label{table: mass of Xi-bb and Xi-bc}
\begin{tabular}{ccccccccccc}\hline
&Ref. \cite{Brown:2014ena}&Ref. \cite{Weng:2018mmf}&Ref. \cite{Ebert:2002ig}&Ref. \cite{Roberts:2007ni}&Ref. \cite{Albertus:2006ya}&Ref. \cite{SilvestreBrac:1996bg}&Refs. \cite{Karliner:2014gca,Karliner:2018hos}&Ref. \cite{He:2004px}&Ref. \cite{Weng:2010rb}& Ref. \cite{Wang:2010hs,Wang:2010vn}\\
\hline
$\Xi_{bb}$&10143(30)(23)&10168.9$\pm$9.2&10202&10340&10197$_{-17}^{+10}$&10204&10162$\pm$12&10272&10090$\pm$10&10170$\pm$140\\
$\Xi_{bb}^{*}$&10178(30)(24)&10188$\pm$7.1&10237&10367&10236$_{-17}^{+9}$&-&10184$\pm$12&-&10337&10220$\pm$150\\
$\Omega_{bb}$&10273(27)(20)&10259.0$\pm$15.5&10359&10454&10260$_{-34}^{+14}$&10258&10208$\pm$18&10369&10180$\pm$5&10320$\pm$140\\
$\Omega_{bb}^{*}$&10308(27)(21)&10267.5$\pm$12.1&10389&10486&10297$_{-28}^{+5}$&-&-&10429&-&10380$\pm$140\\
$\Xi_{bc}$&6943(33)(28)&6922.3$\pm$6.9&6933&7011&6919$_{-7}^{+17}$&6932&6914$\pm$13&6838&6840$\pm$10&-\\
$\Xi_{bc}^{'}$&6959(36)(28)&6947.9$\pm$6.9&6963&7047&6948$_{-6}^{+17}$&-&6933$\pm$12&7028&-&-\\
$\Xi_{bc}^{*}$&6985(36)(28)&6973.2$\pm$5.5&6980&7074&6986$_{-5}^{+14}$&-&6960$\pm$14&6986&-&-\\
$\Omega_{bc}$&6998(27)(20)&7010.7$\pm$9.3&7088&7136&6986$_{-17}^{+27}$&6996&6968$\pm$19&6941&6945$\pm$5&-\\
$\Omega_{bc}^{'}$&7032(28)(20)&7047.0$\pm$9.3&7116&7165&7009$_{-15}^{+24}$&-&6984$\pm$19&7116&-&-\\
$\Omega_{bc}^{*}$&7059(28)(21)&7065.7$\pm$7.5&7130&7187&7046$_{-9}^{+11}$&-&-&7077&-&-\\
\hline
\end{tabular}
\end{table}

In the literature, numerous analyses on the masses of $\Xi_{bb}$ and $\Xi_{bb}^*$ have been performed (see Table I of Ref. \cite{Wei:2016jyk} for a collection). We list several of the results in Table \ref{table: mass of Xi-bb and Xi-bc}, where the involved approaches include lattice QCD \cite{Brown:2014ena}, chromomagnetic models \cite{Weng:2018mmf,Karliner:2014gca,Karliner:2018hos}, relativistic quark model \cite{Ebert:2002ig}, nonrelativistic quark model \cite{Roberts:2007ni,Albertus:2006ya,SilvestreBrac:1996bg}, bag model \cite{He:2004px}, and Bethe-Salpeter equation \cite{Weng:2010rb}. To select an appropriate value for the mass of $\Xi_{bb}$, we adopt the following criteria: 1) the baryon masses satisfy the light-flavor symmetry $\Omega_{bb}^{*}-\Omega_{bb}=\Xi_{bb}^{*}-\Xi_{bb}$ in the heavy quark limit; 2) the HDAS relation $\Omega_{bb}^{*}-\Xi_{bb}^{*}=\bar{B}_{s}^*-\bar{B}^*\approx$91 MeV holds; and 3) the inequality $\Xi_{bb}-(\Xi_{bb})_{CMI}-2(\bar{B}-\bar{B}_{CMI})<\Xi_{cc}-(\Xi_{cc})_{CMI}-2(D-D_{CMI})$ for the mass of $\Xi_{bb}$ is required. The final criterion leading to $\Xi_{bb}<10327$ MeV means that the color-Coulomb contribution to the bottom diquark is larger than that of the charm case once the contributions from the effective quark masses and color-magnetic interactions have been subtracted. It can be confirmed that a similar inequality holds for heavy quarkonia, $\eta_{c}-(\eta_{c})_{CMI}-2(D-D_{CMI})>\eta_{b}-(\eta_{b})_{CMI}-2(\bar{B}-\bar{B}_{CMI})$ (numerically, $-882$ MeV $>$ $-1181$ MeV). The lattice results in Ref. \cite{Brown:2014ena} meet the first and third criteria but not the second because $\Omega_{bb}^{*}-\Xi_{bb}^{*}\approx 130\text{ MeV}>91$ MeV. In the chromomagnetic models \cite{Weng:2018mmf,Karliner:2014gca,Karliner:2018hos}, the results are compatible with all the criteria. In this case, we use the ground baryon mass $\Xi_{bb}=10169$ MeV from Ref. \cite{Weng:2018mmf}, whereas the mass of $\Xi_{bb}^*$ is evaluated to be $\Xi_{bb}^*=\Xi_{bb}+16C_{bn}=10190$ MeV. The mass of $\Omega_{bb}^*$ is, subsequently, further determined to be $\Omega_{bb}^{*}=\Xi_{bb}^*+\bar{B}_s^*-\bar{B}^*=10280$ MeV and that of $\Omega_{bb}$ is $\Omega_{bb}^{*}-16C_{bs}=10266$ MeV.

\begin{figure}[!ht]
\subfigure[$\,\,bb\bar{n}\bar{n}$]{ \label{fig:bbnn}
\includegraphics[width=130pt]{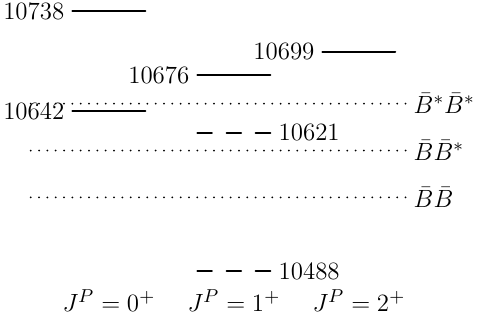}}\hspace{20pt}
\subfigure[$\,\,bb\bar{n}\bar{s}$]{ \label{fig:bbns}
\includegraphics[width=130pt]{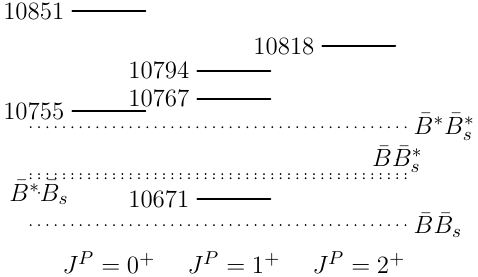}}\hspace{20pt}
\subfigure[$\,\,bb\bar{s}\bar{s}$]{ \label{fig:bbss}
\includegraphics[width=130pt]{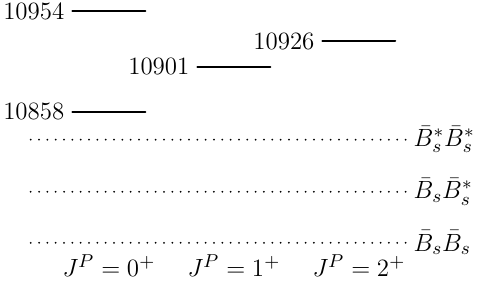}
}
\caption{Relative positions for double-bottom tetraquark states (solid and dashed lines) and relevant meson-meson thresholds (dotted lines). The masses are given in MeV. In (a), the solid (dashed) lines denote the $I=1$ ($I=0$) states.}\label{fig:bb}
\end{figure}

By repeating the procedure for studying the double-charm tetraquarks, we similarly obtain the masses of the highest-spin double-bottom tetraquark states,
\begin{eqnarray}
T_{bb\bar{n}\bar{n}}^{I=1,J=2}&=&\Xi_{bb}^*+\Sigma_b^*-\bar{B}^*=10699\text{ MeV},\nonumber\\
T_{bb\bar{n}\bar{s}}^{J=2}&=&\Xi_{bb}^*+\Xi_b^*-\bar{B}^*=10818\text{ MeV},\nonumber\\
T_{bb\bar{s}\bar{s}}^{J=2}&=&\Omega_{bb}^*+\Omega_{b}^*-\bar{B}_s^*=10926\text{ MeV}.
\end{eqnarray}
These values are 49 MeV, 78 MeV, and 95 MeV (increasing numbers) higher than the $\bar{B}^*\bar{B}^*$, $\bar{B}^*\bar{B}_s^*$, and $\bar{B}_s^*\bar{B}_s^*$ thresholds, respectively. Such mass distances from the corresponding thresholds in Ref. \cite{Luo:2017eub} are approximately 45 MeV, 29 MeV, and 13 MeV (decreasing numbers). With the newly obtained masses of the spin-2 tetraquarks, we can further estimate those of the lower double-bottom states in the CMI model. We list all the results in Table \ref{table:bbQQ} and plot the relative positions for the $bb\bar{q}\bar{q}$ tetraquarks in Fig. \ref{fig:bb}.

\begin{table}[!h]
\caption{Results for the $bb\bar{q}\bar{q}$ ($q=u,d,s$) states (unit: MeV). The second and third columns provide the numerical values of the CMI matrices and their eigenvalues, respectively. The fifth and sixth columns list the masses estimated using Eq. \eqref{eq:ref} (with $\bar{B}^{(*)}\bar{B}^{(*)}$/$\bar{B}^{(*)}\bar{B}_{s}^{(*)}$/$\bar{B}_{s}^{(*)}\bar{B}_{s}^{(*)}$ as the reference state) and Eq. \eqref{modelhamiltonian} (parameters provided in Table \ref{CMI-parameters}), respectively. They can be viewed as the theoretical lower limits (low.) and upper limits (up.), respectively, for the tetraquark masses in the current framework. The fourth column displays our predictions with the heavy diquark-antiquark symmetry (HDAS) consideration.}\label{table:bbQQ}
\begin{tabular}{c|ccccc}\hline
$(bb\bar{n}\bar{n})$&$\langle H_{CM}\rangle$&Eigenvalues&Mass (our)&Mass (low.)&Mass (up.)\\
$I(J^{P})$&&&HDAS&$\bar{B}^{(*)}\bar{B}^{(*)}$&Eq. \eqref{modelhamiltonian}\\
\hline
$1(2^{+})$&$\left(\begin{array}{c}64.7\\\end{array}\right)$&$\left(\begin{array}{c}64.7\\\end{array}\right)$&$\left(\begin{array}{c}10698.7\\\end{array}\right)$&$\left(\begin{array}{c}10691.3\\\end{array}\right)$&$\left(\begin{array}{c}10897.1\\\end{array}\right)$\\
$1(1^{+})$&$\left(\begin{array}{c}42.3\\\end{array}\right)$&$\left(\begin{array}{c}42.3\\\end{array}\right)$&$\left(\begin{array}{c}10676.3\\\end{array}\right)$&$\left(\begin{array}{c}10668.9\\\end{array}\right)$&$\left(\begin{array}{c}10874.7\\\end{array}\right)$\\
$1(0^{+})$&$\left(\begin{array}{cc}31.1&41.2\\41.2&80.3\\\end{array}\right)$&$\left(\begin{array}{c}7.8\\103.7\\\end{array}\right)$&$\left(\begin{array}{c}10641.7\\10737.6\\\end{array}\right)$&$\left(\begin{array}{c}10634.3\\10730.2\\\end{array}\right)$&$\left(\begin{array}{c}10840.2\\10936.1\\\end{array}\right)$\\
$0(1^{+})$&$\left(\begin{array}{cc}-141.7&-23.8\\-23.8&-17.3\\\end{array}\right)$&$\left(\begin{array}{c}-146.1\\-12.9\\\end{array}\right)$&$\left(\begin{array}{c}10487.9\\10621.0\\\end{array}\right)$&$\left(\begin{array}{c}10480.5\\10613.6\\\end{array}\right)$&$\left(\begin{array}{c}10686.3\\10819.5\\\end{array}\right)$\\
\hline
\hline
$(bb\bar{n}\bar{s})$&$\langle H_{CM}\rangle$&Eigenvalues&Mass (our)&Mass (low.)&Mass (up.)\\
$J^{P}$&&&HDAS&$\bar{B}^{(*)}\bar{B}_{s}^{(*)}$&Eq. \eqref{modelhamiltonian}\\
\hline
$2^{+}$&$\left(\begin{array}{c}48.5\\\end{array}\right)$&$\left(\begin{array}{c}48.5\\\end{array}\right)$&$\left(\begin{array}{c}10817.6\\\end{array}\right)$&$\left(\begin{array}{c}10764.7\\\end{array}\right)$&$\left(\begin{array}{c}11061.5\\\end{array}\right)$\\
$1^{+}$&$\left(\begin{array}{ccc}25.0&0.0&0.0\\0.0&-91.3&-24.9\\0.0&-24.9&-8.9\\\end{array}\right)$&$\left(\begin{array}{c}-98.2\\-2.0\\25.0\\\end{array}\right)$&$\left(\begin{array}{c}10670.9\\10767.2\\10794.1\\\end{array}\right)$&$\left(\begin{array}{c}10618.0\\10714.2\\10741.2\\\end{array}\right)$&$\left(\begin{array}{c}10914.8\\11011.0\\11038.0\\\end{array}\right)$\\
$0^{+}$&$\left(\begin{array}{cc}13.3&43.1\\43.1&55.1\\\end{array}\right)$&$\left(\begin{array}{c}-13.7\\82.1\\\end{array}\right)$&$\left(\begin{array}{c}10755.4\\10851.2\\\end{array}\right)$&$\left(\begin{array}{c}10702.5\\10798.3\\\end{array}\right)$&$\left(\begin{array}{c}10999.3\\11095.1\\\end{array}\right)$\\
\hline
\hline
$(bb\bar{s}\bar{s})$&$\langle H_{CM}\rangle$&Eigenvalues&Mass (our)&Mass (low.)&Mass (up.)\\
$J^{P}$&&&HDAS&$\bar{B}_{s}^{(*)}\bar{B}_{s}^{(*)}$&Eq. \eqref{modelhamiltonian}\\
\hline
$2^{+}$&$\left(\begin{array}{c}32.2\\\end{array}\right)$&$\left(\begin{array}{c}32.2\\\end{array}\right)$&$\left(\begin{array}{c}10925.6\\\end{array}\right)$&$\left(\begin{array}{c}10838.1\\\end{array}\right)$&$\left(\begin{array}{c}11225.8\\\end{array}\right)$\\
$1^{+}$&$\left(\begin{array}{c}7.7\\\end{array}\right)$&$\left(\begin{array}{c}7.7\\\end{array}\right)$&$\left(\begin{array}{c}10901.0\\\end{array}\right)$&$\left(\begin{array}{c}10813.6\\\end{array}\right)$&$\left(\begin{array}{c}11201.3\\\end{array}\right)$\\
$0^{+}$&$\left(\begin{array}{cc}-4.6&45.1\\45.1&29.9\\\end{array}\right)$&$\left(\begin{array}{c}-35.6\\60.9\\\end{array}\right)$&$\left(\begin{array}{c}10857.7\\10954.3\\\end{array}\right)$&$\left(\begin{array}{c}10770.3\\10866.8\\\end{array}\right)$&$\left(\begin{array}{c}11158.0\\11254.5\\\end{array}\right)$\\
\hline
\end{tabular}
\end{table}

By comparing the current results with those in Ref. \cite{Luo:2017eub}, a similar $bb\bar{n}\bar{n}$ spectrum can be found. According to Fig. \ref{fig:bbnn}, the $T_{bb}$ state with a mass of $10488$ MeV is approximately 116 MeV below the $\bar{B}\bar{B}^*$ threshold (10604 MeV), and it should be rather stable, but other $bb\bar{n}\bar{n}$ states are not. This observation is the same as that in Ref. \cite{Luo:2017eub}. For the $bb\bar{n}\bar{s}$ states, the masses in the current work are approximately 50 MeV higher than those in Ref. \cite{Luo:2017eub}. According to Fig. \ref{fig:bbns}, only the lowest $T_{bb\bar{n}\bar{s}}^{J=1}$, which is slightly ($\sim$20 MeV) below the $\bar{B}\bar{B}_s^*$ threshold, is possibly a stable tetraquark. This conclusion is similar to that of Ref. \cite{Luo:2017eub}. For the heaviest $bb\bar{s}\bar{s}$ states, the masses in the current work are approximately 80 MeV higher than those in Ref. \cite{Luo:2017eub}, and no stable state is found in both approaches. Therefore, the HDAS and threshold approaches yield similar conclusions regarding the state stabilities for double-bottom tetraquarks.

It is interesting that the masses of the $bb\bar{n}\bar{n}$ states from HDAS consideration coincide with those from the threshold approach. This coincidence may mean that the $(bb)(\bar{n}\bar{n})$ diquark-antidiquark structure and the $(b\bar{n})(b\bar{n})$ molecule-like structure have similar effects on the mass spectrum. It probably also implies that the four quark components have an almost equal spatial distance. If this is true, it is also possible that more than one structure exists near the $\bar{B}\bar{B}^*$ threshold \cite{Yu:2019sxx}. Of course, future experimental data are required to evaluate which approach provides more reasonable results and how large the effects from mass uncertainty of $\Xi_{bb}$ would be.\\

HDAS relations similar to Eqs. \eqref{susyrelation-nn}, \eqref{susyrelation-ns}, and \eqref{susyrelation-ss} can also be applied to bottom-charm tetraquark systems with the replacement $cc\to bc$. It should be noted that there are two bases for the highest-spin $bc\bar{n}\bar{n}$ and $bc\bar{n}\bar{s}$ tetraquarks. For the former case, the color-triplet (color-sextet) diquark exists only in the isovector (isoscalar) state. For the latter case, the two bases are nearly uncoupled, and the color-triplet diquark exists mainly in the higher state. The color-sextet contributions to the higher $T_{bc\bar{n}\bar{s}}^{J=2}$ are not a concern.

\begin{figure}[htbp]
\subfigure[$\,\,bc\bar{n}\bar{n}$]{ \label{fig: bcnn}
\includegraphics[width=130pt]{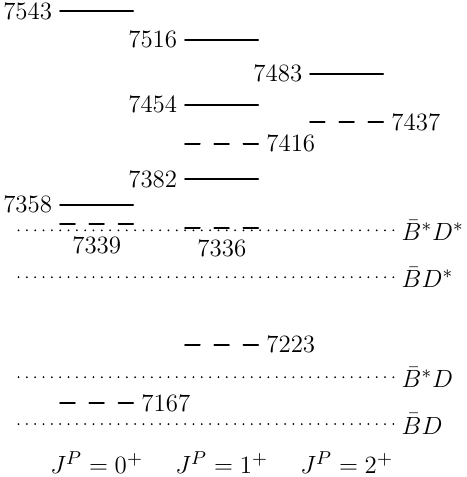}}\hspace{20pt}
\subfigure[$\,\,bc\bar{n}\bar{s}$]{ \label{fig: bcns}
\includegraphics[width=130pt]{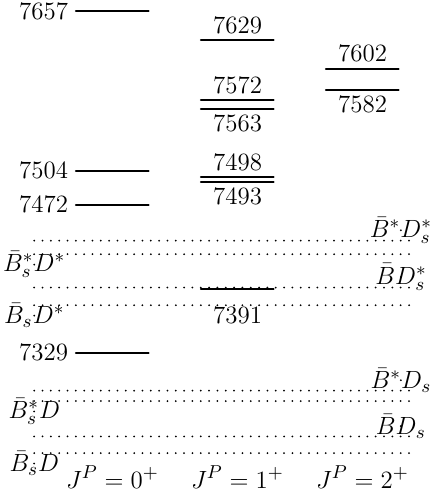}}\hspace{20pt}
\subfigure[$\,\,bc\bar{s}\bar{s}$]{ \label{fig: bcss}
\includegraphics[width=130pt]{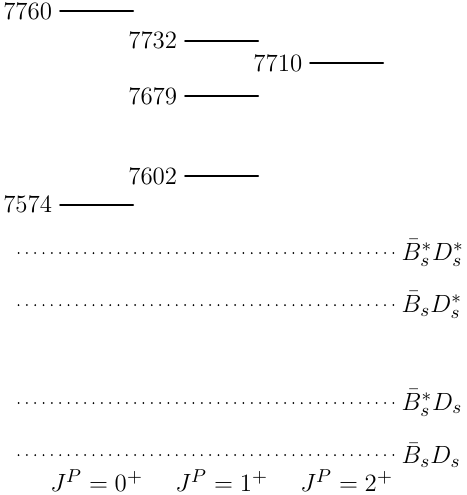}}
\caption{Relative positions for bottom-charm tetraquark states (solid and dashed lines) and relevant meson-meson thresholds (dotted lines). The masses are given in MeV. In (a), the solid (dashed) lines denote the $I=1$ ($I=0$) states.}\label{fig:bc}
\end{figure}

Using the mass $\Xi_{bc}=6922.3\pm6.9$ MeV from Ref. \cite{Weng:2018mmf} and our CMI model, we obtain $\Xi_{bc}^*=6974$ MeV \cite{Liu:2019zoy} and $\Omega_{bc}^*=\Xi_{bc}^*+\bar{B}_s^*-\bar{B}^*=7065$ MeV. Thereafter,
\begin{eqnarray}\label{J2:bcqq}
T_{bc\bar{n}\bar{n}}^{I=1,J=2}&=&\Xi_{bc}^*+\Sigma_b^*-\bar{B}^*=7483\text{ MeV},\nonumber\\
T_{bc\bar{n}\bar{s}}^{J=2,higher}&=&\Xi_{bc}^*+\Xi_b^*-\bar{B}^*=7602\text{ MeV},\nonumber\\
T_{bc\bar{s}\bar{s}}^{J=2}&=&\Omega_{bc}^*+\Omega_{b}^*-\bar{B}_s^*=7710\text{ MeV},
\end{eqnarray}
are obtained. These values are 150 MeV, 165 MeV, and 183 MeV (increasing numbers) higher than the $\bar{B}^*D^*$, $\bar{B}^*D_s^*$, and $\bar{B}_s^*D_s^*$ thresholds, respectively. In Ref. \cite{Luo:2017eub}, such states are approximately 34 MeV, 5 MeV, and 1 MeV (decreasing numbers) higher than the corresponding thresholds. With the reference scales in Eq. \eqref{J2:bcqq}, we obtain the numerical results for the $bc\bar{q}\bar{q}$ tetraquark states in the CMI model. These are listed in Table \ref{table:Tbcqq}, and the spectra are plotted in Fig. \ref{fig:bc}.

\begin{table}[htbp]
\caption{Results for the $bc\bar{q}\bar{q}$ ($q=u,d,s$) states (unit: MeV). The second and third columns provide the numerical values of the CMI matrices and their eigenvalues, respectively. The fifth and sixth columns list the masses estimated using Eq. \eqref{eq:ref} (with $\bar{B}^{(*)}D^{(*)}$/$\bar{B}^{(*)}D_{s}^{(*)}$/$\bar{B}_{s}^{(*)}D_{s}^{(*)}$ as the reference state) and Eq. \eqref{modelhamiltonian} (parameters provided in Table \ref{CMI-parameters}), respectively. These can be viewed as the theoretical lower limits (low.) and upper limits (up.), respectively, for the tetraquark masses in the current framework. The fourth column displays our predictions with the heavy diquark-antiquark symmetry (HDAS) consideration.}\label{table:Tbcqq}
\begin{tabular}{c|ccccc}\hline
$(bc\bar{n}\bar{n})$&$\langle H_{CM}\rangle$&Eigenvalues&Mass (our)&Mass (low.)&Mass (up.)\\
$I(J^P)$&&&HDAS&$\bar{B}^{(*)}D^{(*)}$&Eq. \eqref{modelhamiltonian}\\
\hline
$1(2^{+})$&$\left(\begin{array}{c}77.4\\\end{array}\right)$&$\left(\begin{array}{c}77.4\\\end{array}\right)$&$\left(\begin{array}{c}7483.2\\\end{array}\right)$&$\left(\begin{array}{c}7363.8\\\end{array}\right)$&$\left(\begin{array}{c}7579.5\\\end{array}\right)$\\
$1(1^{+})$&$\left(\begin{array}{ccc}31.0&17.0&36.0\\17.0&32.6&-49.2\\36.0&-49.2&70.5\\\end{array}\right)$&$\left(\begin{array}{c}-24.0\\47.9\\110.2\\\end{array}\right)$&$\left(\begin{array}{c}7381.9\\7453.7\\7516.0\\\end{array}\right)$&$\left(\begin{array}{c}7262.5\\7334.3\\7396.6\\\end{array}\right)$&$\left(\begin{array}{c}7478.1\\7550.0\\7612.3\\\end{array}\right)$\\
$1(0^{+})$&$\left(\begin{array}{cc}7.8&85.2\\85.2&81.3\\\end{array}\right)$&$\left(\begin{array}{c}-48.3\\137.4\\\end{array}\right)$&$\left(\begin{array}{c}7357.5\\7543.2\\\end{array}\right)$&$\left(\begin{array}{c}7238.1\\7423.8\\\end{array}\right)$&$\left(\begin{array}{c}7453.8\\7639.5\\\end{array}\right)$\\
\hline
$0(2^{+})$&$\left(\begin{array}{c}30.9\\\end{array}\right)$&$\left(\begin{array}{c}30.9\\\end{array}\right)$&$\left(\begin{array}{c}7436.7\\\end{array}\right)$&$\left(\begin{array}{c}7317.3\\\end{array}\right)$&$\left(\begin{array}{c}7533.0\\\end{array}\right)$\\
$0(1^{+})$&$\left(\begin{array}{ccc}-85.1&36.0&42.4\\36.0&-141.0&-49.2\\42.4&-49.2&-16.3\\\end{array}\right)$&$\left(\begin{array}{c}-182.7\\-70.2\\10.5\\\end{array}\right)$&$\left(\begin{array}{c}7223.1\\7335.6\\7416.3\\\end{array}\right)$&$\left(\begin{array}{c}7103.7\\7216.2\\7296.9\\\end{array}\right)$&$\left(\begin{array}{c}7319.4\\7431.9\\7512.6\\\end{array}\right)$\\
$0(0^{+})$&$\left(\begin{array}{cc}-143.1&85.2\\85.2&-162.6\\\end{array}\right)$&$\left(\begin{array}{c}-238.6\\-67.0\\\end{array}\right)$&$\left(\begin{array}{c}7167.2\\7338.8\\\end{array}\right)$&$\left(\begin{array}{c}7047.8\\7219.4\\\end{array}\right)$&$\left(\begin{array}{c}7263.5\\7435.1\\\end{array}\right)$\\
\hline
\hline
$(bc\bar{n}\bar{s})$&$\langle H_{CM}\rangle$&Eigenvalues&Mass (our)&Mass (low.)&Mass (up.)\\
$J^{P}$&&&HDAS&$\bar{B}^{(*)}D_{s}^{(*)}$&Eq. \eqref{modelhamiltonian}\\
\hline
$2^{+}$&$\left(\begin{array}{cc}61.0&0.3\\0.3&40.3\\\end{array}\right)$&$\left(\begin{array}{c}40.3\\61.0\\\end{array}\right)$&$\left(\begin{array}{c}7581.5\\7602.2\\\end{array}\right)$&$\left(\begin{array}{c}7429.8\\7450.5\\\end{array}\right)$&$\left(\begin{array}{c}7723.0\\7743.7\\\end{array}\right)$\\
$1^{+}$&$\left(\begin{array}{cccccc}13.8&-0.6&16.8&-0.3&35.6&-1.2\\-0.6&-90.6&-0.1&35.6&0.0&-50.1\\16.8&-0.1&15.8&-1.2&-50.1&0.0\\-0.3&35.6&-1.2&-77.7&-1.4&42.0\\35.6&0.0&-50.1&-1.4&45.3&-0.3\\-1.2&-50.1&0.0&42.0&-0.3&-7.9\\\end{array}\right)$&$\left(\begin{array}{c}-150.3\\-48.2\\-43.5\\21.9\\30.6\\88.1\\\end{array}\right)$&$\left(\begin{array}{c}7390.9\\7493.0\\7497.7\\7563.1\\7571.8\\7629.3\\\end{array}\right)$&$\left(\begin{array}{c}7239.2\\7341.4\\7346.1\\7411.5\\7420.2\\7477.7\\\end{array}\right)$&$\left(\begin{array}{c}7532.4\\7634.5\\7639.2\\7704.6\\7713.3\\7770.8\\\end{array}\right)$\\
$0^{+}$&$\left(\begin{array}{cccc}-9.8&0.2&-0.6&86.7\\0.2&-112.2&86.7&0.0\\-0.6&86.7&-136.7&0.6\\86.7&0.0&0.6&56.1\\\end{array}\right)$&$\left(\begin{array}{c}-212.0\\-69.6\\-36.8\\115.9\\\end{array}\right)$&$\left(\begin{array}{c}7329.2\\7471.6\\7504.3\\7657.1\\\end{array}\right)$&$\left(\begin{array}{c}7177.5\\7319.9\\7352.7\\7505.4\\\end{array}\right)$&$\left(\begin{array}{c}7470.7\\7613.1\\7645.9\\7798.6\\\end{array}\right)$\\
\hline
\hline
$(bc\bar{s}\bar{s})$&$\langle H_{CM}\rangle$&Eigenvalues&Mass (our)&Mass (low.)&Mass (up.)\\
$J^{P}$&&&HDAS&$\bar{B}_{s}^{(*)}D_{s}^{(*)}$&Eq. \eqref{modelhamiltonian}\\
\hline
$2^{+}$&$\left(\begin{array}{c}44.6\\\end{array}\right)$&$\left(\begin{array}{c}44.6\\\end{array}\right)$&$\left(\begin{array}{c}7710.1\\\end{array}\right)$&$\left(\begin{array}{c}7523.9\\\end{array}\right)$&$\left(\begin{array}{c}7907.9\\\end{array}\right)$\\
$1^{+}$&$\left(\begin{array}{ccc}-3.4&16.6&35.2\\16.6&-1.0&-50.9\\35.2&-50.9&20.1\\\end{array}\right)$&$\left(\begin{array}{c}-63.7\\13.0\\66.4\\\end{array}\right)$&$\left(\begin{array}{c}7601.9\\7678.5\\7731.9\\\end{array}\right)$&$\left(\begin{array}{c}7415.6\\7492.3\\7545.7\\\end{array}\right)$&$\left(\begin{array}{c}7799.6\\7876.3\\7929.7\\\end{array}\right)$\\
$0^{+}$&$\left(\begin{array}{cc}-27.4&88.2\\88.2&30.9\\\end{array}\right)$&$\left(\begin{array}{c}-91.1\\94.6\\\end{array}\right)$&$\left(\begin{array}{c}7574.4\\7760.1\\\end{array}\right)$&$\left(\begin{array}{c}7388.1\\7573.9\\\end{array}\right)$&$\left(\begin{array}{c}7772.2\\7957.9\\\end{array}\right)$\\
\hline
\end{tabular}
\end{table}

For the $bc\bar{n}\bar{n}$ system, stable states are not found from our results. However, if the errors in the adopted approach are considered, the lowest $I(J^P)=0(0^+)$ and $0(1^+)$ tetraquarks may be around the $\bar{B}D$ and $\bar{B}^*D$ thresholds, respectively. This conclusion differs from that in Ref. \cite{Luo:2017eub}, where these two states and the isoscalar spin-2 state are all stable. In Ref. \cite{Cheng:2020nho}, we investigated the $bc\bar{n}\bar{n}$ spectrum with a reference scale related to the $X(4140)$ by assuming it to be a $cs\bar{c}\bar{s}$ tetraquark. It is interesting that the masses of the $bc\bar{n}\bar{n}$ states in that approach are consistent with the present results. The conclusion that the lowest $0^+$ $T_{bc}$ may be around the $\bar{B}D$ threshold is also consistent with the findings in Ref. \cite{Karliner:2017qjm}. For the $(bc\bar{n}\bar{s})$ and $(bc\bar{s}\bar{s})$ systems, according to Fig. \ref{fig:bc}, no stable tetraquarks can be found, which is consistent with the conclusion in Ref. \cite{Cheng:2020nho} but different from that in Ref. \cite{Luo:2017eub}, where stable $bc\bar{n}\bar{s}$ is still possible. Although the masses of $bc\bar{n}\bar{n}$ agree with those in Ref. \cite{Cheng:2020nho}, those of the $bc\bar{n}\bar{s}$ and $bc\bar{s}\bar{s}$ states are higher. Future experiments will be required to evaluate which approach, threshold, X(4140), or HDAS, is better.

\section{Constraints on masses of $QQq$, $QQQ$, and $QQ\bar{q}\bar{q}$ states}\label{sec5}

We have obtained the masses of the double-heavy tetraquark states with the aid of heavy diquark-antiquark symmetry. The values are all larger than those in the threshold approach in Ref. \cite{Luo:2017eub}, in which reference meson-meson thresholds were adopted. For the $bc\bar{n}\bar{n}$ and $bc\bar{n}\bar{s}$ states, the current masses are also heavier than those in the $X(4140)$ approach in Ref. \cite{Cheng:2020nho}, in which the reference scales were related to the mass of $X(4140)$. At present, it cannot be judged where such states are actually located, as there is still no observed double-heavy tetraquark. Even for conventional baryons, a state containing two heavy quarks that are heavier than $\Xi_{cc}$ has not been reported. It is helpful to make a theoretical estimation on the range of their masses to understand the possible structures of an observed state. In Tables \ref{table:Tcc coupled channel}, \ref{table:Tccpartners}, \ref{table:bbQQ}, and \ref{table:Tbcqq}, we treat the masses obtained using Eq. \eqref{modelhamiltonian} as the upper limits. In fact, the range may be narrowed further from another perspective. We now investigate the constraints on the masses of the $QQq$ and $QQQ$ baryons first, and then those of the double-heavy tetraquarks. If the experiments reveal a state with a larger mass than the obtained limit, that hadron should not be a ground state with high probability.

Suppose that we are estimating the mass of a multiquark state $X$ using Eq. \eqref{eq:ref}. In principle, several reference hadron-hadron systems can be adopted. Their thresholds will result in different values for the mass of the $X$ state. Considering two reference systems $\mathcal{A+D}$ and $\mathcal{B+C}$, where $\mathcal{A,B,C,D}$ are four hadrons, and $\mathcal{A+D}$ and $\mathcal{B+C}$ have the same quark content as $X$, two values for the mass of $X$ ($M_{\mathcal{AD}}$ and $M_{\mathcal{BC}}$) can be obtained. An inequality between these must exist. For convenience, we further assume that they satisfy $M_{\mathcal{AD}}< M_{\mathcal{BC}}$. Thus, according to the estimation formula \eqref{eq:ref}, we obtain
\begin{eqnarray}\label{mass-constraint}
[M_ \mathcal{A}-\langle H_{CMI}\rangle_{\mathcal{A}}]+[M_\mathcal{D}-\langle H_{CMI}\rangle_{\mathcal{D}}] < [M_\mathcal{B}-\langle H_{CMI}\rangle_{\mathcal{B}}]+[M_\mathcal{C}-\langle H_{CMI}\rangle_{\mathcal{C}}],
\end{eqnarray}
where $M_{\cal A,B,C,D}$ should be the measured masses and no longer the theoretical masses obtained using Eq. \eqref{modelhamiltonian} or \eqref{eq:ref}. This formula means that the color-electric interactions in the hadrons have different effects on the two reference hadron-hadron systems. In many cases, it has been found that systems with $M_{\cal A}>M_{\cal B}/M_{\cal C}>M_{\cal D}$ satisfy the inequality \cite{Wu:2016gas,Zhou:2018pcv,Wu:2017weo}. If the assignment for the four hadrons can be provided, and the masses of three of them have been measured, a constraint on the mass of the fourth hadron will be obtained. In this case, we do not demand that the multiquark state $X$ must exist. For the involved states in the present study, ${\cal A}$ has two heavy quarks, whereas ${\cal B}/{\cal C}$ has one heavy quark. Because the binding between two heavy quarks owing to the color-Coulomb potential is positively associated with their reduced mass, the $QQ$ attraction inside ${\cal A}$ is large. When the quark structure in ${\cal A}+{\cal D}$ is changed to ${\cal B}+{\cal C}$, the two heavy quarks require additional energy so that they can be separated and recombined with other light quarks into two single-heavy hadrons. Subsequently, an inequality (\ref{mass-constraint}) naturally follows. It should be noted that it is different from the Hall-Post inequalities, which link $N$-body energies to $N^\prime$-body energies ($N^\prime<N$) \cite{Richard:2019cmi}.

As an example, we consider the ground $\Xi_{cc}$ state. When we discuss the case whereby ${\cal A}=\Xi_{cc}$ and ${\cal D}$ is a light meson, we can select the following four sets of reference meson-baryon systems,
$$
\text{I}:\,\Xi_{cc}+\pi\leftrightarrow\Sigma_{c}+D;\quad \text{II}:\, \Xi_{cc}+K\leftrightarrow\Sigma_{c}+D_{s}; \quad
\text{III}:\, \Xi_{cc}+\bar{K}\leftrightarrow\Xi_{c}^\prime+D; \quad \text{IV}:\, \Xi_{cc}+\phi\leftrightarrow\Xi_{c}^\prime+D_{s}.
$$
Their corresponding $X$ states are $ccnn\bar{n}$, $ccnn\bar{s}$, $ccns\bar{n}$, and $ccns\bar{s}$, respectively. When we adopt the inequality $M_{\cal AD}< M_{\cal BC}$, we can obtain the upper limits for the mass of $\Xi_{cc}$: 3770 MeV, 3694 MeV, 3735 MeV, and 3663 MeV, respectively. Their difference may be a considerable 100 MeV, but the lowest value 3663 MeV should be used as the constraint. This value is approximately 40 MeV larger than the LHCb result $\Xi_{cc}^{++}=3621$ MeV \cite{Aaij:2017ueg}. To obtain the mass constraint, we can also use higher spin states herein, such as $\Xi_{cc}+\rho\rightarrow\Sigma_{c}+D^{*}$, but no new results are obtained. Another case that can be discussed is that in which ${\cal D}$ is a light baryon. Five sets of reference baryon-baryon systems can be considered:
$$
\text{I}:\Xi_{cc}+N\leftrightarrow\Sigma_{c}+\Sigma_{c};\,\,\text{II}:\Xi_{cc}+\Sigma\leftrightarrow\Sigma_{c}+\Xi_{c}^\prime; \,\,
\text{III}:\Xi_{cc}+\Xi\leftrightarrow\Xi_{c}^\prime+\Xi_{c}^\prime; \,\,
\text{IV}:\Xi_{cc}+\Xi\leftrightarrow\Sigma_{c}+\Omega_{c}; \,\, \text{V}:\Xi_{cc}+\Omega\leftrightarrow\Xi_{c}^\prime+\Omega_{c}.
$$
Their corresponding $X$ states are $ccnnnn$, $ccnnns$, $ccnnss$, $ccnnss$, and $ccnsss$, respectively. The upper limits that are otained are 3776 MeV, 3732 MeV, 3717 MeV, 3709 MeV, and 3657 MeV, respectively. The situation is very similar to the above case. Combining the analyses in these two cases, we obtain $\Xi_{cc}<3657$ MeV, which is approximately 30 MeV higher than the measured value.

Extending the discussions to $\Xi_{bb}$ and $\Xi_{bc}$, we can similarly determine the minimum upper limits for their masses. We display the relevant reference states and results in Table \ref{table: doubly heavy baryon up limit}. The obtained constraints are $\Xi_{bb}<10319$ MeV and $\Xi_{bc}<6972$ MeV. Similar constraints can also be found in Ref. \cite{Liu:2019zoy}. In fact, more stringent constraints are possible.

\begin{table}[htbp]
\caption{$\mathcal{D}\mathcal{B}\mathcal{C}$ states in constraining the upper limits for the masses of the $QQn$ baryons and the obtained limits (Up.) in MeV. Here, ``Meson'' (``Baryon'') means that $\mathcal{D}$ is a light-quark meson (baryon).}\label{table: doubly heavy baryon up limit}
\begin{tabular}{c|ccccccccc}\hline
$\Xi_{cc}$&Meson&$\pi\Sigma_{c}D$&$K\Sigma_{c}D_{s}$&$K\Xi_{c}^{'}D$&$\phi\Xi_{c}^{'}D_{s}$&&&&\\
&Up.&3770&3694&3735&3663&&&&\\
&Baryon&$N\Sigma_{c}\Sigma_{c}$&$\Sigma\Sigma_{c}\Xi_{c}^{'}$&$\Xi\Xi_{c}^{'}\Xi_{c}^{'}$&$\Xi\Sigma_{c}\Omega_{c}$&$\Omega\Xi_{c}^{'}\Omega_{c}$&&&\\
&Up.&3775&3732&3717&3709&3657&&&\\
\hline
$\Xi_{bb}$&Meson&$\pi\Sigma_{b}\bar{B}$&$K\Sigma_{b}\bar{B}_{s}$&$K\Xi_{b}^{'}\bar{B}$&$\phi\Xi_{b}^{'}\bar{B}_{s}$&&&&\\
&Up.&10466&10377&10423&10339&&&&\\
&Baryon&$N\Sigma_{b}\Sigma_{b}$&$\Sigma\Sigma_{b}\Xi_{b}^{'}$&$\Xi\Xi_{b}^{'}\Xi_{b}^{'}$&$\Xi\Sigma_{b}\Omega_{b}$&$\Omega\Xi_{b}^{'}\Omega_{b}$&&&\\
&Up.&10462&10412&10389&10379&10319&&&\\
\hline
$\Xi_{bc}$&Meson&$\pi\Sigma_{b}D$&$\pi\Sigma_{c}B$&$K\Sigma_{b}D_{s}$&$K\Sigma_{c}\bar{B}_{s}$&$\phi\Xi_{b}^{'}D_{s}$&$\phi\Xi_{c}^{'}\bar{B}_{s}$&&\\
&Up.&7102&7111&7027&7022&6988&6991&&\\
&Baryon&$N\Sigma_{b}\Sigma_{c}$&$\Sigma\Sigma_{b}\Xi_{c}^{'}$&$\Sigma\Sigma_{c}\Xi_{b}^{'}$&$\Xi\Sigma_{b}\Omega_{c}$&$\Xi\Sigma_{c}\Omega_{b}$&$\Xi\Xi_{b}^{'}\Xi_{c}^{'}$&$\Omega\Xi_{b}^{'}\Omega_{c}$&$\Omega\Xi_{c}^{'}\Omega_{b}$\\
&Up.&7108&7065&7057&7042&7024&7042&6982&6972\\
\hline
\end{tabular}
\end{table}

In Sec. \ref{subsec: Spectrum of Tbb and Tbc system}, we obtained the constraint $\Xi_{bb}<10327$ MeV with the inequality $\Xi_{bb}-(\Xi_{bb})_{CMI}-2(\bar{B}-\bar{B}_{CMI})<\Xi_{cc}-(\Xi_{cc})_{CMI}-2(D-D_{CMI})$ while determining an appropriate mass of $\Xi_{bb}$. This inequality, similar to \eqref{mass-constraint}, also arises from the color-Coulomb interaction between two heavy quarks. Naturally, the mass of $\Xi_{bc}$ can also be taken into consideration, and we obtain
\begin{eqnarray}\label{mass-Xicc-Xibc-Xibb in meson}
&&\Xi_{bb}-(\Xi_{bb})_{CMI}-2(\bar{B}-\bar{B}_{CMI})\nonumber\\
&<&\Xi_{bc}-(\Xi_{bc})_{CMI}-(\bar{B}-\bar{B}_{CMI})-(D-D_{CMI})\notag\\
&<&\Xi_{cc}-(\Xi_{cc})_{CMI}-2(D-D_{CMI}).
\end{eqnarray}
The constraint $\Xi_{bc}<6963$ MeV subsequently follows. Replacing the reference mesons with reference baryons, we similarly obtain
\begin{eqnarray}\label{mass-Xicc-Xibc-Xibb in baryon}
&&\Xi_{bb}-(\Xi_{bb})_{CMI}-2(\Sigma_{b}-(\Sigma_{b})_{CMI})\nonumber\\
&<&\Xi_{bc}-(\Xi_{bc})_{CMI}-(\Sigma_{b}-(\Sigma_{b})_{CMI})-(\Sigma_{c}-(\Sigma_{c})_{CMI})\notag\\
&<&\Xi_{cc}-(\Xi_{cc})_{CMI}-2(\Sigma_{c}-(\Sigma_{c})_{CMI}).
\end{eqnarray}
Now, slightly smaller numbers ($\Xi_{bb}<10308$ MeV and $\Xi_{bc}<6954$ MeV) than those shown in Table \ref{table: doubly heavy baryon up limit} are obtained.

According to the above discussions, our short summary on the mass constraints for the doubly heavy baryons is as follows: $\Xi_{cc}<3657$ MeV, $\Xi_{bb}<10308$ MeV, and $\Xi_{bc}<6954$ MeV. Of course, the constraint on $\Xi_{bb}$ may be updated once the mass of $\Xi_{bc}$ is measured, or vice versa. If an observed $\Xi_{bb}$ or $\Xi_{bc}$ has a larger mass than the limit provided here, it should not be the ground state. In the previous sections, the masses adopted were $\Xi_{cc}=3621$ MeV, $\Xi_{bb}=10169$ MeV, and $\Xi_{bc}=6922$ MeV, which satisfy the obtained constraints. We do not discuss the $QQs$ case. At present, without experimental data regarding the $QQs$ baryons, we cannot obtain more stringent constraints than those provided in Ref. \cite{Liu:2019zoy}.

Using the same concept, we can estimate the upper limits for the masses of triply heavy baryons. These rely on the masses of doubly heavy baryons. We use $\Xi_{cc}=3621$ MeV, $\Xi_{bc}<6954$ MeV, and $\Xi_{bb}<10308$ MeV as the inputs and list our results for all of the ground triply heavy baryons in Table \ref{table: Triply heavy baryon up limit}, from which $\Omega_{ccc}<4962$ MeV, $\Omega_{ccb}<8250$ MeV, $\Omega_{bbc}<11578$ MeV, and $\Omega_{bbb}<14939$ MeV can be obtained. Similar constraints can also be found in Ref. \cite{Liu:2019zoy}. The upper limits for the $\Omega_{ccb}$, $\Omega_{bbc}$, and $\Omega_{bbb}$ states could be changed to lower values if the masses of $\Xi_{bc}$ and $\Xi_{bb}$ are measured experimentally.
\begin{table}[htbp]
\caption{$\mathcal{D}\mathcal{B}\mathcal{C}$ states in constraining the upper limits for the masses of the $QQQ$ baryons and the obtained limits (Up.) in MeV. Here, ``Meson'' (``Baryon'') means that $\mathcal{D}$ is a light-quark meson (baryon).}\label{table: Triply heavy baryon up limit}
\begin{tabular}{c|ccccccc}\hline
$\Omega_{ccc}$&Meson&$\pi\Xi_{cc}D$&$K\Xi_{cc}D_{s}$&&&&\\
&Up.&5038&4962&&&&\\
&Baryon&$N\Xi_{cc}\Sigma_{c}$&$\Sigma\Xi_{cc}\Xi_{c}^{'}$&$\Xi\Xi_{cc}\Omega_{c}$&&&\\
&Up.&5043&5000&4977&&&\\
\hline
$\Omega_{ccb}$&Meson&$\pi\Xi_{cc}B$&$K\Xi_{cc}B_{s}$&$\pi\Xi_{bc}D$&$K\Xi_{bc}D_{s}$&&\\
&Up.&8339&8250&8298&8254&&\\
&Baryon&$N\Xi_{cc}\Sigma_{b}$&$\Sigma\Xi_{cc}\Xi_{b}^{'}$&$\Xi\Xi_{cc}\Omega_{b}$&$N\Xi_{bc}\Sigma_{c}$&$\Sigma\Xi_{bc}\Xi_{c}^{'}$&$\Xi\Xi_{bc}\Omega_{c}$\\
&Up.&8336&8285&8252&8336&8292&8269\\
\hline
$\Omega_{bbc}$&Meson&$\pi\Xi_{bc}\bar{B}$&$K\Xi_{bc}\bar{B}_{s}$&$\pi\Xi_{bb}D$&$K\Xi_{bb}D_{s}$&&\\
&Up.&11666&11578&11657&11581&&\\
&Baryon&$N\Xi_{bc}\Sigma_{b}$&$\Sigma\Xi_{bc}\Xi_{b}^{'}$&$\Xi\Xi_{bc}\Omega_{b}$&$N\Xi_{bb}\Sigma_{c}$&$\Sigma\Xi_{bb}\Xi_{c}^{'}$\\
&Up.&11663&11612&11579&11662&11619&11596\\
\hline
$\Omega_{bbb}$&Meson&$\pi\Xi_{bb}\bar{B}$&$K\Xi_{bb}\bar{B}_{s}$&&&&\\
&Up.&16028&14939&&&&\\
&Baryon&$N\Xi_{bb}\Sigma_{b}$&$\Sigma\Xi_{bb}\Xi_{b}^{'}$&$\Xi\Xi_{bb}\Omega_{b}$&&&\\
&Up.&15025&14974&14941&&&\\
\hline
\end{tabular}
\end{table}

Now, we investigate the $QQ\bar{q}\bar{q}$ case. When we consider a double-heavy tetraquark state in terms of the diquark-antiquark symmetry, its mass is linearly dependent on the mass of a related double-heavy baryon, which is treated as an input, see, e.g., Eq. \eqref{susyrelation-nn}. If we use $\Delta^{up.}$ to denote the difference between the upper limit for the mass of this $QQq$ baryon and the mass that we adopt, we can set the upper limit for the mass of any tetraquark state by adding $\Delta^{up.}$ to the obtained tetraquark mass, so that the HDAS relation still holds. Explicitly, we need to add 36 MeV, 139 MeV, and 32 MeV for $cc\bar{q}\bar{q}$, $bb\bar{q}\bar{q}$, and $bc\bar{q}\bar{q}$, respectively. For the lowest $QQ\bar{n}\bar{n}$ tetraquarks, we have $T_{cc}< 3965$ MeV, $T_{bb}< 10627$ MeV ($\approx BB^*$ threshold+17 MeV), and $T_{bc}< 7199$ MeV. However, because the symmetry relations are only approximately correct, the measured tetraquark masses in future experiments may exceed such limits.

Let us return to the upper limits using the inequality \eqref{mass-constraint}. Naturally, the involved $X$ systems are $QQq\bar{q}\bar{q}\bar{q}$ and $QQqqq\bar{q}\bar{q}$. In the latter case, tetraquarks are always involved in the reference channels (${\cal A}+{\cal D}$ or ${\cal B}+{\cal C}$), and we cannot obtain useful information, at least presently. In the former case, we, unfortunately, cannot obtain reliable constraints, either. If $X=ccs\bar{n}\bar{n}\bar{s}$, for example, the reference system can be $(cc\bar{n}\bar{n})(s\bar{s})$, $(cs\bar{n}\bar{n})(c\bar{s})$, or $(ccs)(\bar{n}\bar{n}\bar{s})$. However, in constraining the mass of $cc\bar{n}\bar{n}$, neither $(cs\bar{n}\bar{n})(c\bar{s})$ nor $(ccs)(\bar{n}\bar{n}\bar{s})$ can be adopted. The former system involves another tetraquark state, whereas the latter does not meet the requirement to use the inequality \eqref{mass-constraint} that the two heavy quarks should be separated into two hadrons, which guarantees the difference caused by the color-Coulomb potential. Another reason is that the mass of $ccs$ baryon has not been measured. If we neglect the requirement to use \eqref{mass-constraint} and consider the case where $X=ccn\bar{n}\bar{n}\bar{n}$, the reference state $(ccn)(\bar{n}\bar{n}\bar{n})$ can be adopted and a mass constraint $T_{cc}<3952$ MeV is obtained. This value appears to be the upper limit of the mass, but this is simply conjecture and not a conclusion. Therefore, we could not get more information from the case that ${\cal D}$ is a light hadron. One may wonder whether we can estimate the lower limit of the $QQ\bar{q}\bar{q}$ mass with the minimum theoretical mass of $QQQ$ baryon by considering the case $X=QQQ\bar{q}\bar{q}\bar{q}$. In fact, obtaining the limit is possible, but the constraint is probably not useful. For example, one could get $T_{cc}>3704$ MeV with $\Omega_{ccc}=4790$ MeV \cite{Hasenfratz:1980ka}. This value is approximately 70 MeV smaller than the lower limit given in Table \ref{table:Tcc coupled channel}, and no meaningful constraint is obtained.

Similar to \eqref{mass-Xicc-Xibc-Xibb in meson}, another inequality exists for tetraquarks. In this case, we only consider the case without strange quarks. Subsequently, we obtain
\begin{eqnarray}\label{mass-Tcc-Tbc-Tbb in meson}
&&T_{bb}-(T_{bb})_{CMI}-2(\bar{B}-\bar{B}_{CMI})\nonumber\\
&<& T_{bc}-(T_{bc})_{CMI}-(\bar{B}+D-\bar{B}_{CMI}-D_{CMI}) \notag\\
&<& T_{cc}-(T_{cc})_{CMI}-2(D-D_{CMI}),
\end{eqnarray}
which can be employed to verify the results obtained in the previous sections. As $T_{bb}-(T_{bb})_{CMI}-2(\bar{B}-\bar{B}_{CMI})=8$ MeV, $T_{bc}-(T_{bc})_{CMI}-(\bar{B}+D-\bar{B}_{CMI}-D_{CMI})=120$ MeV, and $T_{cc}-(T_{cc})_{CMI}-2(D-D_{CMI})<156$ MeV, the inequalities are certainly satisfied. We can also consider the inequality to be similar to \eqref{mass-Xicc-Xibc-Xibb in baryon}, but the obtained relations do not change. Thus, the inequality \eqref{mass-Tcc-Tbc-Tbb in meson} is sufficient for the purpose of conducting a simple check on the obtained tetraquark masses.

\section{Discussions and summary}\label{sec: Discussions and summary}

It is known that the mass splittings of conventional hadrons are mainly determined by the chromomagnetic interactions. However, while applying the CMI model \eqref{modelhamiltonian} to hadron masses, the deviations from the experimental data may be large (e.g. Table 2 of Ref. \cite{Liu:2019zoy}).  After all, the model is a simplified version of potential quark models. Contributions from color-Coulomb interaction, color confinement, and others are simply effectively absorbed into the masses of the quarks and coupling parameters. In principle, it is unrealistic to determine all the hadron masses with only one set of parameters. In the multiquark case, we tend to adopt a method to compensate for the above effects partially by selecting a suitable reference system, instead of using Eq. \eqref{modelhamiltonian} directly. This appears to be more reasonable than simply taking a set of effective quark masses as input, but the details of the kinematic and dynamic effects may still lead to a significant shift in the spectrum. To fix the deviation, we can take the other effects into account explicitly by sacrificing the concision and
simplicity of calculation. However, we can also balance simplicity and rationality in certain peculiar cases, as in the double-heavy tetraquark systems explored here.

When the $\bar{3}_c$ $QQ$ diquark is regarded as a heavier antiquark $\bar{Q}^\prime$, the double-heavy tetraquark $QQ\bar{q}\bar{q}$ can be viewed as a single-heavy ``antibaryon'' $\bar{Q}^\prime\bar{q}\bar{q}$ in the sense that they have the same color configuration. The mass relations among the $QQ\bar{q}\bar{q}$, $QQq$, $Qqq$, and $Q\bar{q}$ states follow such a heavy diquark-antiquark symmetry. As only the highest-spin $QQ\bar{q}\bar{q}$ states may contain the pure $\bar{3}_c$ $QQ$ diquark, and no mixing effects are involved, their masses are determined with the symmetry relations, and they are selected as the reference states to obtain the tetraquark spectra. Another consideration for using the highest-spin states is that the spin-dependent terms between the light quarks will be cancelled, and those between the heavy and light quarks can be ignored.

\begin{table}[!h]
\caption{Stability of the double-heavy tetraquarks in various studies. The meanings of ``S'', ``US'', and ``ND'' are ``stable'', ``unstable'', and ``not determined'', respectively.}\label{table: stability of various articles}
\begin{tabular}{cccccccccc}\hline
Reference&$(cc\bar{n}\bar{n})$&$(cc\bar{n}\bar{s})$&$(cc\bar{s}\bar{s})$&$(bb\bar{n}\bar{n})$&$(bb\bar{n}\bar{s})$&$(bb\bar{s}\bar{s})$&$(bc\bar{n}\bar{n})$&$(bc\bar{n}\bar{s})$&$(bc\bar{s}\bar{s})$\\
\hline
This work&US&US&US&S&S&US&ND&US&US\\
\cite{Lee:2009rt}&S&S&&S&S&&S&US&\\
\cite{Luo:2017eub}&S&S&US&S&S&US&S&S&US\\
\cite{Pepin:1996id}&S&&&S&&&&&\\
\cite{Feng:2013kea}&S&&&S&&&S&&\\
\cite{Karliner:2017qjm}&US&&&S&&&S&&\\ 
\cite{Park:2018wjk}&US&&&S&S&&US&US&\\
\cite{Yang:2019itm}&S&&&S&&&S&&\\
\cite{Deng:2018kly}&S&US&US&S&S&US&S&US&US\\
\cite{Tan:2020ldi}&S&&&S&&&S&&\\
\cite{Lu:2020rog}&US&US&US&S&US&US&US&US&US\\
\cite{Ebert:2007rn}&US&US&US&S&US&US&US&US&US\\
\cite{Yang:2020fou}&&&US&&&US&&&US\\
\cite{Du:2012wp}&US&US&US&S&S&S&&&\\ 
\cite{Chen:2013aba}&&&&&&&&S&S\\
\cite{Navarra:2007yw}&US&&&S&&&&&\\
\cite{Francis:2016hui,Francis:2018jyb}&US&US&&S&S&&S&US&\\

\cite{Leskovec:2019ioa}&&&&&&&S&&\\ 
\cite{Hudspith:2020tdf}&&&&S&S&&US&US&\\
\cite{Carlson:1987hh}&US&&&S&&&ND&&\\ 
\cite{Cheng:2020nho}&&&&&&&ND&US&\\

\cite{Eichten:2017ffp}&US&US&US&S&S&US&US&US&US\\

\cite{Braaten:2020nwp}&US&US&US&S&S&US&US&US&US\\ 
\hline
\end{tabular}
\end{table}

Once the masses of all the double-heavy tetraquarks are obtained, it is easy to determine whether or not stable tetraquarks exist from Figs. \ref{fig:cc}-\ref{fig:bc}. In Table \ref{table: stability of various articles}, we present our answers to the question. In fact, numerous discussions on double-heavy tetraquarks can be found in the literature \cite{Liu:2019zoy}. For example, Carlson et. \cite{Carlson:1987hh} discussed non-strange $QQ\bar{q}\bar{q}$ systems and found that $T_{bb}$ is sufficiently stable against strong decay, $T_{cc}$ is unstable, and $T_{bc}$ is uncertain. For comparison, we have also displayed the results obtained in some reference studies in the table. In general, all the studies support the stable double-bottom tetraquark $T_{bb}$. The results indicate that the double-charm $T_{cc}$ state is probably unstable, whereas the stability of $T_{bc}$ remains controversial.

The consistency between our results and others indicates that the estimation method with HDAS is reasonable. However, how reliable the numerical results are is not clear because they are affected by several factors. First, the accuracy of the approximate HDAS relations and errors of the input $QQq$ masses determine the location of the tetraquark spectra. Second, the existence of $(QQ)_{6_c}(\bar{q}\bar{q})_{\bar{6}_c}$ configuration may significantly affect the mass splittings if the color-electric contributions are considered explicitly. Third, the values of $C_{ij}$ determining the mass splittings are extracted from conventional hadrons. Whether they can be applied to multiquark states remains an open question.

The spacial structure of the tetraquark states was not considered in the above discussions. An observed double-heavy state can also be a meson-meson molecule, the spatial structure of which differs from the compact tetraquark. At present, it is generally difficult to determine a criterion to distinguish a compact multiquark state from a molecular state, but there are cases where this is possible. In the $bb\bar{n}\bar{n}$ case, both compact tetraquark and molecules \cite{Ohkoda:2012hv,Li:2012ss,Xu:2017tsr,Wang:2018atz,Sakai:2017avl} are possible, but the binding energies in these two configurations differ. It is possible to identify the inner structure of the observed state: a large (small) binding energy corresponds to a compact (molecular) state. However, in the $cc\bar{n}\bar{n}$ case, the observed state should be a molecule \cite{Ohkoda:2012hv,Li:2012ss,Xu:2017tsr} if it is below the related meson-meson threshold. In the $bc\bar{n}\bar{n}$ case, both molecules \cite{Li:2012ss,Sakai:2017avl} and compact tetraquarks are around the related meson-meson thresholds. The situation is complicated, and further discussions are required.

\begin{table}[!htbp]
\caption{Strong and electromagnetic decay patterns for the lowest tetraquark states.}\label{table:Decay}
\setlength{\tabcolsep}{7mm}{
\begin{tabular}{ccccc}\hline
System&Mass&\multicolumn{2}{c}{Strong Decay}& Electromagnetic Decay\\
&(MeV)&2 body&3 body& 3 body \\
\hline
$(cc\bar{n}\bar{n})^{J=1}_{I=0}$&3929&$DD^{*}$&$DD\pi$&$DD\gamma$/$DD^{*}\gamma$\\
$(cc\bar{n}\bar{s})^{J=1}$&4106&$D^{*}D_{s}$/$DD^{*}_{s}$&$DD_{s}\pi$&$DD_{s}\gamma$/$D^{*}D_{s}\gamma$/$DD^{*}_{s}\gamma$\\
$(cc\bar{s}\bar{s})^{J=0}$&4227&$D_{s}D_{s}$&$D_{s}D_{s}^{*}\pi$&$D_{s}D_{s}\gamma$/$D_{s}D^{*}_{s}\gamma$\\
$(bb\bar{n}\bar{n})^{J=1}_{I=0}$&10488&---&---&---\\
$(bb\bar{n}\bar{s})^{J=1}$&10671&---&---&$\bar{B}\bar{B}_{s}\gamma$\\
$(bb\bar{s}\bar{s})^{J=0}$&10858&$\bar{B}_{s}\bar{B}_{s}$/$\bar{B}_{s}^{*}\bar{B}_{s}^{*}$&---&$\bar{B}_{s}\bar{B}_{s}\gamma/\bar{B}_{s}\bar{B}^{*}_{s}\gamma/\bar{B}^{*}_{s}\bar{B}^{*}_{s}\gamma$\\
$(bc\bar{n}\bar{n})^{J=0}_{I=0}$&7167&$\bar{B}D$&---&$\bar{B}D\gamma$\\
$(bc\bar{n}\bar{s})^{J=0}$&7329&$\bar{B}_{s}D/\bar{B}D_{s}$&---&$\bar{B}_{s}D\gamma$/$\bar{B}D_{s}\gamma$/$\bar{B}^{*}_{s}D\gamma$/$\bar{B}^{*}D_{s}\gamma$\\
$(bc\bar{s}\bar{s})^{J=0}$&7574&$\bar{B}_{s}D_{s}/\bar{B}^{*}_{s}D^{*}_{s}$&$\bar{B}^{*}_{s}D_{s}\pi$&$\bar{B}_{s}D_{s}\gamma$/$\bar{B}^{*}_{s}D_{s}\gamma$/$\bar{B}_{s}D^{*}_{s}\gamma$/$\bar{B}^{*}_{s}D^{*}_{s}\gamma$\\
\hline
\end{tabular}
}
\end{table}

The detailed partial widths for the studied tetraquarks depend on the Hamiltonian and specific processes, and a quantitative calculation will be discussed in future work. Here, we present a brief analysis on their dominant decay patterns. In Table \ref{table:Decay}, the strong and electromagnetic decay patterns for the lowest state in each system are provided. For higher states, we simply mention the rearrangement decay modes. The thresholds of such meson-meson channels are illustrated in Figs. \ref{fig:cc}-\ref{fig:bc}. Whether or not the decays can occur is determined mainly by the kinematics and quantum number conservations. In the $cc\bar{n}\bar{n}$ case, the allowed channels for the $I(J^P)=0(1^+)$ state are $DD^*$ and $D^*D^*$, those for the $1(0^+)$ states are $DD$ and $D^*D^*$, those for the $1(1^+)$ state are $DD^*$ and $D^*D^*$, whereas $DD$, $DD^*$, and $D^*D^*$ are all allowed channels for the $1(2^+)$ state. The channels in the case $bb\bar{n}\bar{n}$ are similar. In the $cc\bar{s}\bar{s}$ and $bb\bar{s}\bar{s}$ cases, the channels can be obtained with reference to the $I=1$ cases. In the $cc\bar{n}\bar{s}$ case, the $0^+$ states can decay into $DD_s$ and $D^*D_s^*$, the $1^+$ states can decay into $DD_s^*/D^*D_s$ and $D^*D_s^*$, and the $2^+$ state can decay into all these channels. The case of $bb\bar{n}\bar{s}$ is similar. In the case of $bc\bar{n}\bar{n}$, the allowed channels for the $0^+$ states are $\bar{B}D$ and $\bar{B}^*D^*$, those for the $1^+$ states are $\bar{B}^*D/\bar{B}D^*$ and $\bar{B}^*D^*$, and all these decay channels are allowed for the $2^+$ state. The channels in the case of $bc\bar{s}\bar{s}$ can be obtained with the replacement $n\to s$. In the case of $bc\bar{n}\bar{s}$, the allowed decay channels for the $0^+$ states are $\bar{B}_sD/\bar{B}D_s$ and $\bar{B}^*D_s^*/\bar{B}_s^*D^*$, those for the $1^+$ states are $\bar{B}_s^*D/\bar{B}^*D_s$ and $\bar{B}^*_sD^*/B^*D_s^*$, and those for the $2^+$ states are $\bar{B}_sD/\bar{B}D_s$, $\bar{B}_s^*D/\bar{B}^*D_s$ and $\bar{B}^*_sD^*/B^*D_s^*$.

If the studied compact double-heavy tetraquarks exist, one may wonder where and how to search for them according to the decay channels. In principle, they can be produced at any collider if the collision energy is sufficiently high. For example, they may be produced in the $Z$ boson decay \cite{Ali:2018ifm}, hadron decays \cite{Esposito:2013fma}, $pp$ collision \cite{Ali:2018xfq}, heavy-ion collisions \cite{Cho:2010db,Cho:2011ew,Hong:2018mpk,Fontoura:2019opw}, and $e^+e^-$ annihilation process \cite{Hyodo:2012pm,Hyodo:2017hue,Jin:2014nva}. A low production rate and small signal/noise ratio should be the main reasons that double-heavy tetraquarks are not observed. Owing to the clean background, the $e^+e^-$ annihilation process offers its own advantage in searching for a double-charm state. As a signal of double-charm tetraquarks has not been observed in such a process, the detection efficiency should be increased with improved analysis methods, such as that proposed in Ref. \cite{Jin:2014nva}.

In summary, we have determined the masses of the highest-spin double-heavy tetraquark states $QQ\bar{q}\bar{q}$ with the aid of the heavy diquark-antiquark symmetry. Thereafter, such reference states were used to derive the masses of their partners with mass splittings in the CMI model. We presented the results in Tables \ref{table:Tcc coupled channel}, \ref{table:Tccpartners}, \ref{table:bbQQ}, and \ref{table:Tbcqq} as well as Figs. \ref{fig:cc}-\ref{fig:bc}. The double-charm tetraquarks $cc\bar{q}\bar{q}$ ($q=u,d,s$) were significantly higher than their rearrangement decay channels, and we did not find a bound state in the systems. In the double-bottom systems $bb\bar{q}\bar{q}$, we did not obtain stable $bb\bar{s}\bar{s}$ states, but observed a deep $bb\bar{n}\bar{n}$ bound state ($T_{bb}$) and a shallow $bb\bar{n}\bar{s}$ bound state. Their $I(J^P)$ were both $0(1^+)$, and their masses were 10488 MeV ($\approx116$ MeV below the $\bar{B}\bar{B}^*$ threshold) and 10671 MeV ($\approx20$ MeV below the $\bar{B}\bar{B}_s^*/\bar{B}_s\bar{B}^*$ threshold), respectively. The $T_{bb}$ mass is very close to that obtained in Ref. \cite{Luo:2017eub} when $\bar{B}\bar{B}^*$ was used as the reference state. For the bottom-charm systems $(bc\bar{q}\bar{q})$, no stable $bc\bar{n}\bar{s}$ or $bc\bar{s}\bar{s}$ was found, but we could obtain two near-threshold $bc\bar{n}\bar{n}$ states. These were the lowest $0(0^+)$ state $T_{bc}$, with a mass of 7167 MeV, and the lowest $0(1^{+})$ state, with a mass of 7223 MeV. Considering the model uncertainties, it was difficult to draw a conclusion whether or not they were stable. From our results, it can be concluded that the order of possibility for finding a bound $QQ\bar{q}\bar{q}$ tetraquark should be $(cc)<(bc)<(bb)$ and that a bound tetraquark becomes more difficult to form with the increasing number of strange quarks. Because not all of the input masses were measured, we also discussed the constraints on the masses of heavy quark hadrons, such as $\Xi_{bb}<10308$ MeV and $\Omega_{ccc}<4962$ MeV. We obtained $T_{cc}<3965$ MeV, $T_{bb}<10627$ MeV, and $T_{bc}<7199$ MeV for the lowest tetraquark states. Of course, whether or not this is true requires future experimental tests. We hope that our predictions for double-heavy states will be helpful for future investigations.

\section*{Acknowledgments}\label{sec: Acknowledgments}

This project is supported by National Natural Science Foundation of China (Grant Nos. 11775130, 11775132,
11635009, 11325525, 11875179) and by Natural Science Foundation of Shandong Province (Grant Nos. ZR2016AM16, ZR2017MA002).


\end{document}